\documentclass[sigconf]{acmart}

\usepackage{color}
\usepackage{pifont}
\usepackage{subfigure} 
\usepackage{float}

\usepackage{booktabs}
\usepackage{multirow}

\usepackage[toc,page]{appendix} 
\usepackage{tabularx}
\usepackage{url}
\usepackage{enumitem}
\setlist[itemize]{leftmargin=*}

\newcommand{\ms}[1]{{\color{black} #1}}

\usepackage{listings}
\usepackage[ruled]{algorithm2e}

\AtBeginDocument{%
  \providecommand\BibTeX{{%
    \normalfont B\kern-0.5em{\scshape i\kern-0.25em b}\kern-0.8em\TeX}}}

\copyrightyear{2025}
\acmYear{2025}
\setcopyright{acmlicensed}\acmConference[CHI '25]{CHI Conference on Human Factors in Computing Systems}{April 26-May 1, 2025}{Yokohama, Japan}
\acmBooktitle{CHI Conference on Human Factors in Computing Systems (CHI '25), April 26-May 1, 2025, Yokohama, Japan}
\acmDOI{10.1145/3706598.3713423}
\acmISBN{979-8-4007-1394-1/25/04}

\begin{document}




\title{Towards Human-AI Deliberation: Design and Evaluation of LLM-Empowered Deliberative AI for AI-Assisted Decision-Making}





\author{Shuai Ma}
\orcid{0000-0002-7658-292X}
\affiliation{
  \institution{The Hong Kong University of Science and Technology}
  \city{Hong Kong}
  \country{China}
}
\email{shuai.ma@connect.ust.hk}

\author{Qiaoyi Chen}
\affiliation{
  \institution{The Hong Kong University of Science and Technology}
  \city{Hong Kong}
  \country{China}
}
\email{qchench@connect.ust.hk}

\author{Xinru Wang}
\affiliation{
  \institution{Purdue University}
  \city{West Lafayette}
  \country{USA}
}
\email{xinruw@purdue.edu}

\author{Chengbo Zheng}
\affiliation{
  \institution{The Hong Kong University of Science and Technology}
  \city{Hong Kong}
  \country{China}
}
\email{cb.zheng@connect.ust.hk}

\author{Zhenhui Peng}
\affiliation{
  \institution{Sun Yat-sen University}
  \city{Zhuhai}
  \country{China}
}
\email{pengzhh29@mail.sysu.edu.cn}

\author{Ming Yin}
\affiliation{
  \institution{Purdue University}
  \city{West Lafayette}
  \country{USA}
}
\email{mingyin@purdue.edu}

\author{Xiaojuan Ma}
\affiliation{
  \institution{The Hong Kong University of Science and Technology}
  \city{Hong Kong}
  \country{China}
}
\email{mxj@cse.ust.hk}

\renewcommand{\shortauthors}{Shuai Ma, et al.}

\begin{abstract}

Traditional AI-assisted decision-making systems often provide fixed recommendations that users must either accept or reject entirely, limiting meaningful interaction—especially in cases of disagreement. To address this, we introduce \emph{Human-AI Deliberation}, an approach inspired by human deliberation theories that enables dimension-level opinion elicitation, iterative decision updates, and structured discussions between humans and AI. At the core of this approach is \emph{Deliberative AI}, an assistant powered by large language models (LLMs) that facilitates flexible, conversational interactions and precise information exchange with domain-specific models. Through a mixed-methods user study, we found that \emph{Deliberative AI} outperforms traditional explainable AI (XAI) systems by fostering appropriate human reliance and improving task performance. By analyzing participant perceptions, user experience, and open-ended feedback, we highlight key findings, discuss potential concerns, and explore the broader applicability of this approach for future AI-assisted decision-making systems.

\end{abstract}



\maketitle

\section{Introduction}

With remarkable technological advancements, AI has been increasingly used to support people in making decisions in various domains, including criminal justice \cite{dodge2019explaining, dressel2018accuracy}, admissions \cite{cheng2019explaining, zhang2023deliberating}, financial investment \cite{green2019principles}, and medical diagnosis \cite{cai2019hello, lee2021human}, among others. Concerns surrounding AI's accuracy, safety, ethics, and accountability \cite{cai2019hello, lee2021human, binns2018s} have led to the widespread adoption of the \emph{AI-assisted decision-making} paradigm in real-world applications \cite{buccinca2021trust, zhang2020effect, wang2021explanations, bansal2021does}. In this paradigm, AI performs an assistive role by providing a recommendation, while human decision-makers can choose to accept or reject it in their final decision \cite{lai2021towards}.

\begin{figure*}[htbp]
	\centering 
	\includegraphics[width=1\linewidth]{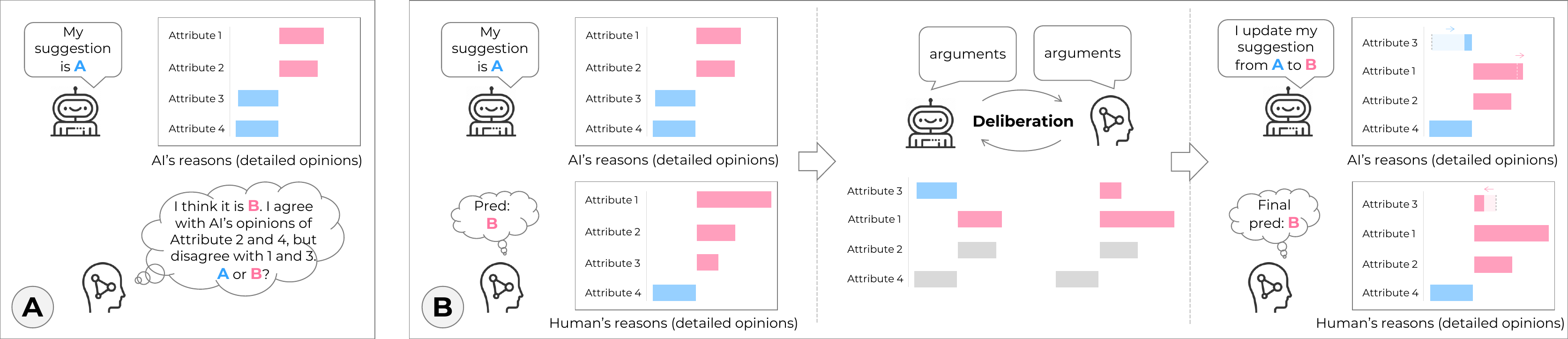}
	\caption{An illustration of \emph{Human-AI Deliberation}. (A) In traditional AI-assisted decision-making, when humans disagree with AI's suggestions (and only find parts of AI's reasons applaudable), it is difficult for humans to decide whether and how much to adopt AI's suggestion. (B) In our proposed \emph{Human-AI Deliberation}, we provide opportunities for the human and the AI model to deliberate on conflicting opinions by discussing related evidence and arguments. Then, AI and humans can update their thoughts (when find it necessary) and reach final predictions.}
	\label{fig:illustration}
        \Description{}
\end{figure*}

Research in recent years, however, identified two challenges within the existing AI-assisted decision-making paradigm.  \textbf{[Challenge 1]} First, a battery of empirical studies found that people rarely trigger analytical thinking when directly presented with AI's suggestions \cite{buccinca2020proxy, rastogi2020deciding, bertrand2022cognitive}. 
As a result, people frequently over-rely on the AI's incorrect recommendations (over-reliance) or mistakenly ignore AI's correct suggestions (under-reliance) \cite{buccinca2021trust, ma2023should, wang2021explanations}. 
Although some solutions have been proposed, such as displaying AI explanations \cite{bansal2021does} and forcing people to think more effortfully \cite{buccinca2021trust}, the results are mixed at best \cite{lai2022human, poursabzi2021manipulating}.
\textbf{[Challenge 2]} Second, instead of full consensus or complete divergence, human and AI decision rationales often exhibit partial alignment \cite{sivaraman2023ignore, wang2023watch}. While they may concur on certain aspects, differences may persist on others \cite{miller2023explainable}. However, in current AI-assisted decision-making systems, AI always provides a fixed recommendation regardless of human thoughts and humans can only accept or reject AI's recommendation \emph{as a whole} \cite{lai2021towards}, with limited support for resolving conflicts or engaging in a meaningful exchange of ideas with the AI system \cite{miller2023explainable}. For example, as shown in Figure \ref{fig:illustration} (a), when the human decision-maker's prediction is inconsistent with the AI model's recommendation and the human only partially agrees with the AI's reasoning (e.g., explanation), existing AI-assisted decision-making interfaces do not support any communication between humans and AI regarding conflicting opinions. This limitation may impede the effective utilization of both human and AI knowledge, hindering collaborative and complementary human-AI team performance.

Deliberation, characterized by thoughtful and reasoned discussion, plays a pivotal role in facilitating constructive discourse and consensus-building across various contexts \cite{steenbergen2003measuring, bachtiger2019mapping}. Deliberation proves effective in facilitating diverse human decision-making tasks, including deliberative politics \cite{black2010methods, habermas2005concluding, thompson2008deliberative}, clinical diagnosis \cite{schaekermann2019understanding, inguaggiato2019moral, preisz2019fast}, criminal justice \cite{devine2001jury, van2019crowdsourcing}, among others. It offers individuals an opportunity to rigorously evaluate different perspectives, including their own, which can potentially address Challenge 1 in AI-assisted decision-making. Moreover, deliberation allows participants to refine their viewpoints through informed discussions about opinion discrepancies \cite{peirce2014illustrations, good1950probability, gough2007weight}. Such a structured process may also enable humans and AI to engage in detailed discussions, potentially mitigating Challenge 2. 
Despite the potential benefits of deliberation, how to design mechanisms to facilitate deliberative conversation between humans and AI and how deliberations influence AI-assisted decision-making remain to be explored.


\ms{In this paper, building on established guidelines for enhancing discourse quality and identifying common ground in human deliberation \cite{steenbergen2003measuring, bachtiger2019mapping, bachtiger2009measuring}, as well as the weight-of-evidence approach in decision-making \cite{alvarez2021human, weed2005weight, baumann2014decision}, we propose a novel solution: \emph{Human-AI Deliberation} for AI-assisted decision-making (Figure \ref{fig:illustration} (b))}. 
Instead of presenting a fixed AI suggestion for humans to accept or reject, our approach encourages humans to externalize their thoughts, enables an interactive deliberation process between humans and AI around the conflicting points of their opinions and rationales, and fosters dynamic, fine-grained updates of humans and AI's decisions. The key component of this approach is \emph{Deliberative AI}, which has the ability to locate viewpoint dissimilarities, stimulate comprehensive deliberation with human decision-makers, and make necessary changes, even compromises, in its own suggestion as the constructive discussion unfolds. To design such an AI assistant, we propose to integrate the strength of domain-specific models (for reliable assistant information generation) and Large Language Models (LLMs, for interactivity and conversation capability). We elaborate on the architecture design of \emph{Human-AI Deliberation} and \emph{Deliberative AI} in Section 3 and demonstrate how to instantiate the architecture in an illustrative task in Section 4.


Since the primary purpose of deliberation is to resolve conflicts between human and AI perspectives, we intentionally selected task cases with notable human-AI disagreements for our user study, drawing from insights in our pilot study. As human-AI deliberation is designed to both resolve conflicts and mitigate inappropriate human reliance on AI, we investigate its impact on task performance and human reliance on AI. Additionally, since deliberation explicitly highlights these conflicts, we are particularly interested in examining its effect on human perceptions of AI and the overall decision-making experience. Specifically, using our proposed concept of \emph{Human-AI Deliberation} as a research probe, we aim to explore the following research questions.

\begin{itemize}
    \item \textbf{RQ1}: How will \emph{Human-AI Deliberation} affect task performance and humans' reliance on AI suggestions?
    \item \textbf{RQ2}: How will \emph{Human-AI Deliberation} affect humans' perceptions of the AI partner and their user experience?
    \item \textbf{RQ3}: How will humans perceive the effectiveness of the proposed \emph{Human-AI Deliberation} and what can be improved for future \emph{Human-AI Deliberation} design?
\end{itemize}

To answer these questions, we conducted an exploratory study using a graduate admissions task. We recruited participants with graduate admissions experience (at least once admitted to a graduate program) on Prolific and asked them to predict an applicant's chance of getting an offer based on the applicant's profile. We compared the proposed \emph{Deliberative AI} with traditional \emph{explainable AI (XAI)} and \emph{human alone} baselines. Our experimental results revealed that \emph{Human-AI Deliberation} has the potential to enhance decision accuracy and promote appropriate reliance on AI recommendations compared to traditional XAI assistants. We conclude by discussing key implications and addressing generalizability concerns based on our design and study findings.


In summary, we make three contributions:

\begin{itemize}
    
    \item \ms{We propose a novel architecture, \emph{Deliberative AI}, designed to enable \emph{Human-AI Deliberation}—a collaborative process where humans and AI deliberate together to resolve conflicting perspectives in decision-making tasks—by seamlessly integrating domain-specific models with large language models (LLMs).}

    \item We demonstrate the instantiation of the \emph{Deliberative AI} in an illustrative task (college graduate admission), including the implementation of different components and interface design.
    \item We conduct an exploratory study to gain an initial empirical understanding of how \emph{Human-AI Deliberation} might impact the decision-making process and how humans would perceive this novel AI assistance. Additionally, we demonstrate its potential to improve decision accuracy and promote appropriate human reliance on AI.
\end{itemize}

\section{Related Work}

\subsection{AI-Assisted Decision-Making: Objectives and Challenges}

Artificial Intelligence (AI) is increasingly used in decision-making across various domains \cite{ma2019smarteye, ma2022glancee, dastin2018amazon, dilsizian2014artificial}. However, AI's real-world applications are not infallible, still far from 100\% accuracy \cite{shi2023retrolens, gao2018implicit, ma2022modeling}. This is especially concerning in high-stakes domains like medicine and criminal justice, leading to ethical and legal complexities \cite{cai2019hello, lee2021human, binns2018s}. To address this, the prevalent paradigm of \emph{AI-assisted decision-making} has emerged, drawing substantial attention in the Human-Computer Interaction (HCI) and AI communities \cite{buccinca2021trust, zhang2020effect, wang2021explanations, bansal2021does}. In this paradigm, AI takes on an assistant role, offering recommendations for human decision-makers to accept or reject in their final decisions \cite{lai2021towards}.

\ms{Research in AI-assisted decision-making spans a range of objectives, including enhancing team performance \cite{zhang2020effect, ma2023should}, promoting decision fairness \cite{chiang2023two, wangeffects}, improving efficacy and efficiency \cite{cheng2019explaining}, fostering understanding of AI \cite{wang2021explanations, cheng2019explaining}, building trust and appropriate reliance on AI \cite{schemmer2023appropriate, he2023knowing}, and enriching subjective user experiences \cite{liao2021human, ma2022modeling}. Among these, a key goal is achieving complementary performance—where the collaborative outcomes of human-AI teams exceed what either humans or AI could achieve independently \cite{bansal2021does}. Despite its importance, recent empirical studies highlight persistent difficulties in reaching this goal \cite{zhang2020effect, bansal2021does, rastogi2020deciding}, driven by two primary challenges.}

One challenge is the underutilization of human and AI domain knowledge \cite{bansal2020optimizing}. Some researchers aim to leverage the complementary aspects of human and AI intelligence by training AI to complement human knowledge \cite{bansal2019updates, wilder2020learning}. Moreover, existing AI-assistant interfaces do not efficiently harness the knowledge of both parties \cite{steyvers2022three}. AI contributes its knowledge to humans by providing recommendations with AI explanations serving as a means of representing its detailed reasoning \cite{lai2021towards}. These explanations could facilitate the collaborative synthesis of human and AI intelligence, allowing them to combine insights into different features for final decisions. However, when conflicting views arise, current interfaces provide limited support for the communication and exchange of human and AI knowledge \cite{liao2021human}. To address this, we propose \emph{Human-AI Deliberation} to resolve conflicts through natural discussions.

The second challenge concerns human reliance on AI suggestions \cite{bansal2021does, zhang2020effect, ma2023should, ma2024beyond, ma2024you}. Achieving complementary performance relies on human decision-makers' ability to judiciously determine when to consider AI recommendations and when to be skeptical \cite{zhang2020effect, buccinca2021trust, rastogi2020deciding}. Both over-reliance, where individuals trust AI excessively \cite{lee2004trust, parasuraman1997humans}, and under-trust, where individuals fail to utilize AI when necessary \cite{lee2004trust}, can lead to adverse outcomes. Successful decision-making requires individuals to decide whether and how to rely on AI recommendations on a case-by-case basis \cite{zhang2020effect, bansal2019beyond, bansal2019updates, bansal2021does, turner2022calibrating}. Current approaches present AI performance indicators, explanations, outputs, and confidence levels to assist humans in making informed decisions. However, existing research has found that when people are provided with a recommendation and passively look at it, they often lack analytical thinking, leading to over-reliance or under-reliance on AI systems \cite{buccinca2020proxy, gajos2022people, kaur2020interpreting, liao2021human}. \emph{Human-AI Deliberation}, as proposed in this paper, encourages a careful evaluation of AI rationales through discussions of conflicts in human and AI opinions. By engaging humans in the deliberation process, it promotes a more comprehensive understanding of AI insights, reducing the potential for both under-reliance and over-reliance.

\subsection{The Role of Deliberation in Human Decision Making}

The meaning of deliberation is \emph{``the act of thinking about or discussing something and deciding carefully''} \cite{deliberation_dict}. It involves considering all relevant individuals as moral agents who must justify their viewpoints and listen to others' reasons \cite{gracia2003ethical}. Rather than seeking consensus, the process aims to enhance individual perspectives by incorporating others' viewpoints, thus increasing decision maturity and wisdom \cite{gracia2003ethical}. The origin of group deliberation can be traced back to public deliberation or deliberative democracy, where citizens convene to discuss policies with potential implications for their lives \cite{simon2008blackwell}. Recent studies on online deliberation have showcased its ability to enhance the accuracy of crowd-working tasks \cite{chen2019cicero, drapeau2016microtalk}, improve perceptions of procedural justice \cite{fan2020digital}, and facilitate consensus-building among participants \cite{lee2020solutionchat, schaekermann2019understanding, van2019crowdsourcing, xie2020chexplain}. Furthermore, deliberation proves effective in facilitating various decision-making tasks, including clinical diagnosis \cite{schaekermann2019understanding, inguaggiato2019moral, preisz2019fast}, criminal justice \cite{devine2001jury, van2019crowdsourcing}, and more.


Effective decision-making is crucial across various domains, and deliberation offers significant advantages. It improves decision quality and problem-solving by enabling comprehensive analysis and evaluation of options, fostering a deeper understanding of issues and outcomes \cite{barabas2004deliberation, landemore2015deliberation}. Deliberative decisions are often wiser and more effective due to their basis in thorough analysis \cite{landemore2012democratic, dijksterhuis2006making}. Additionally, deliberation promotes participation and collaboration, encouraging stakeholder engagement and facilitating communication, which aids in resolving complex issues and ensuring decision acceptance \cite{nabatchi2015public, wojcieszak2010deliberative, fishkin2018democracy}. Finally, it helps mitigate decision biases and enhance fairness by enabling objective analysis and reducing emotional influences \cite{kramer1990pretrial, hochman2015fairness, tetlock2017expert}.

Despite the significance of deliberation in decision-making, there is a dearth of research on its integration into AI-assisted decision-making processes. To address this gap, drawing upon theories and practices in deliberation \cite{black2010methods, habermas2005concluding, thompson2008deliberative, lord2013politics, steiner2005deliberative}, we propose \emph{Human-AI Deliberation} to facilitate human reflection and discussion on conflicting human-AI opinions. Based on this approach, we aim to move a first step towards designing a \emph{Deliberative AI} and investigating its effects on decision processes and outcomes through an exploratory empirical study.

\subsection{Existing Studies on Deliberation in AI-Assisted Decision Making}
Deliberation enhances decision-making by integrating diverse perspectives, improving solution quality, and fostering critical thinking \cite{gracia2003ethical}. It involves analytical reflection and active discussion\cite{gracia2003ethical}, both of which have been explored separately in research on AI-assisted decision making.

To stimulate analytical thinking, different interventions have been designed to encourage deeper engagement with System 2 thinking \cite{kahneman2011thinking}, such as ``cognitive forcing'' techniques that prompt more deliberation. Examples include asking individuals to make independent predictions before receiving AI input \cite{buccinca2021trust, park2019slow, rastogi2020deciding} or using ``slow algorithms'' to reduce reliance on AI. Additionally, providing AI explanations without concrete recommendations \cite{gajos2022people} and using AI-framed questioning \cite{danry2023don} have been shown to enhance critical thinking. However, these approaches may lead to under-reliance and may not fully address differences between human and AI perspectives.

Some studies have explored human-AI dialogues in cooperative games \cite{meta2022human, kramar2022negotiation}, but these were not tailored to decision-making tasks. Recent work has begun integrating discussions into AI-assisted decision-making. For example, Zheng et al. \cite{zheng2023competent} included AI in group decisions for student essay evaluations, though these efforts often rely on Wizard of Oz setups rather than purpose-built AI systems. Similarly, Chiang et al. \cite{chiang2023two} studied collaboration between AI and two humans in recidivism risk assessment but limited the AI’s role to providing suggestions without active discussion. Other efforts, such as Zhang et al. \cite{zhang2023deliberating}, used AI models as tools to facilitate deliberation among organizations but did not address direct deliberation between humans and AI. Perhaps the most relevant work is by Slack et al. \cite{slack2022talktomodel}, who explored dialogue-based AI explanations to handle follow-up user questions and improve understanding. However, the key difference is that we focus on deliberation design and propose \emph{Deliberative AI} which can not only ``explain to users'' but also actively engage users in the deliberative discussions by ``asking or challenging'' the users, aiming to promote people's critical thinking.

Previous work has touched upon the idea of having AI serve deliberation among humans, but to our knowledge, no research has directly facilitated deliberation between humans and AI. Our work takes an initial step toward designing and evaluating human-AI deliberation in decision-making contexts, offering valuable insights into integrating deliberation into AI-assisted decision-making processes.

\section{Deliberative AI for Human-AI Deliberation}

This section introduces deliberation into the decision-making process, focusing on weighing the evidence. Building on the Weight of Evidence (WoE) framework, we propose a \emph{Human-AI Deliberation} architecture and present the design considerations and architecture of \emph{Deliberative AI}, an AI assistant capable of engaging in deliberation with humans.

\subsection{Integrating Deliberation into Decision-Making}
Decision-making involves selecting the best choice from a range of alternatives to achieve a desired outcome \cite{eisenfuhr2011decision}. The process is typically summarized in seven steps\footnote{Seven-step decision making: Step 1: Identify the problem, Step 2: Collect information, Step 3: Identify the alternatives, Step 4: Weigh the evidence, Step 5: Choose from the alternatives, Step 6: Implement action, and Step 7: Evaluate the results.}, with \emph{weighing the evidence} being a critical step due to its direct influence on subsequent decision outcomes \cite{sevensteps, lunenburg2010decision}. We propose conducting \emph{Human-AI Deliberation} during this phase because disagreements often arise between human perspectives and AI suggestions in this phase and these human-AI disagreements can lead to conflicting opinions and divergent outcomes in later steps \cite{lai2021towards}.

\begin{figure*}[htbp]
	\centering 
	\includegraphics[width=\linewidth]{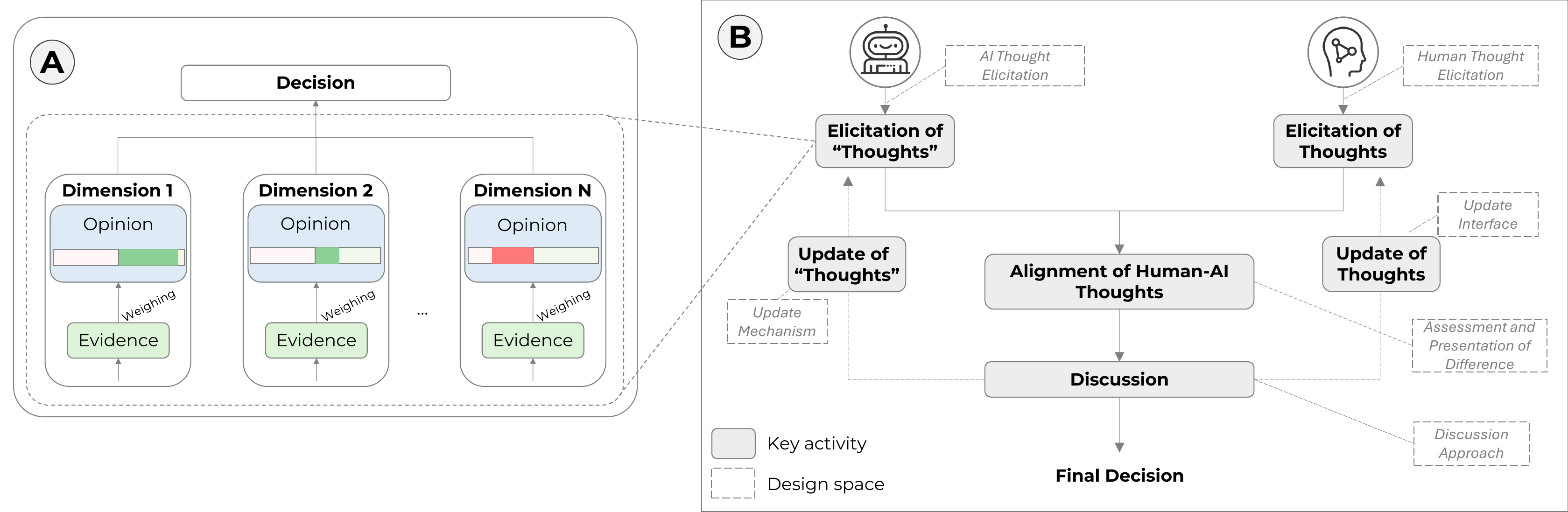}
	\caption{The architecture of \emph{Human-AI Deliberation}. (A) Illustrates the Weight of Evidence (WoE) concept in decision-making, showcasing how decision-makers assess evidence across dimensions to shape opinions and arrive at a final decision. (B) Presents the Architecture for \emph{Human-AI Deliberation}, with key activities (shown in grey boxes) and potential design space (shown in dashed-line boxes).}
	\label{fig:woe_framework}
        \Description{}
\end{figure*}

To structure the deliberation process, we break down decision-making problems and human-AI thoughts into four components (Figure \ref{fig:woe_framework} (a)): (1) \textbf{Decision}: The overall choice to be made for a specific problem; (2) \textbf{Dimension}: An aspect considered when forming the overall decision. In tabular datasets commonly used in decision-making \cite{ghai2021explainable, wang2021explanations}, dimensions often correspond to attributes or sets of related attributes, such as academic excellence or research ability in a graduate admission task; (3) \textbf{Opinion} on a dimension: The assessment of a dimension's impact on the overall decision (e.g., academic excellence contributes +50\% to admission probability); (4) \textbf{Evidence}: The foundation for forming an opinion. For humans, evidence may include facts, heuristics, or personal experiences \cite{simon1990bounded, simon1997models}. For AI, evidence is often rooted in the information encoded in its training data.

We propose deliberation at the dimension level, focusing on how each dimension supports or opposes the overall decision. Both humans and AI must substantiate their opinions with evidence and assess its credibility and probative value \cite{baumann2014decision}. This process generates the Weight of Evidence (WoE) \cite{cartwright2011theory}, which quantifies evidence significance/importance relative to alternatives \cite{peirce2014illustrations, good1950probability, gough2007weight}. WoE is widely used in decision-making for its intuitive meaning and practical implementation \cite{saaty2008decision, gough2007weight}. In the rest of this paper, we will use ``opinion'' and ``WoE'' interchangeably.



\subsection{An Architecture of Human-AI Deliberation}
Based on the WoE-centered decision-making approach, we propose \emph{Human-AI Deliberation}, an architecture to stimulate deliberative processes between humans and AI (Figure \ref{fig:woe_framework}). 
This architecture comprises the following essential activities:
\begin{itemize}
    \item \textbf{Elicitation of Thoughts}: Human and AI start with articulating their dimension-level perspectives on the decision problem. 
    While AI presenting its ``thoughts'' (e.g., in the form of feature importance explanation) is rather common in AI-assisted decision-making \cite{liao2021human}, this activity also encourages individuals to clarify their ideas and examine their reasoning, which prompts analytical thinking in human \cite{milkman2009can, buccinca2021trust}. 
    Two aspects of this activity require careful design. First, AI thought elicitation demands a good balance between human information needs and interpretability \cite{mishra2021crowdsourcing, abdul2020cogam}. Second, human thought elicitation, while encouraging thoughtful reasoning \cite{milkman2009can, buccinca2021trust}, can impose a potential workload. It thus demands suitable, friendly interface designs.
    \item \textbf{Alignment of Human-AI Thoughts}: As human's and AI's viewpoints and the process they form those viewpoints may diverge \cite{kaufman2022cognitive, miller2019explanation, holzinger2019causability}, this activity is tasked with establishing a common language for the two parties to compare their WoE and determine the extent of discrepancy. 
    Proper assessment and presentation of human-AI WoE differences can help effectively navigate humans' attention and efforts in the subsequent activities \cite{boggust2022shared}.   
    \item \textbf{Discussion}: This activity fosters constructive discussions where humans and AI substantiate their opinions, clarify evidence choices, and explain weight assignments. It promotes critical thinking, reduces biases, and highlights differences between parties \cite{milkman2009can, larson1994discussion, hirokawa1985discussion}. With a broad design space, it must be tailored to specific decision tasks, considering factors like content, style, leadership (who initiates and leads the conversation), and duration. A potential solution is adapting human-human discussion \cite{milkman2009can, larson1994discussion, hirokawa1985discussion} to the human-AI discussion contexts.
    \item \textbf{Update of Thoughts}: In-depth discussions may expose potential flaws and conflicts in the original decisions as humans and AI are both imperfect \cite{bansal2021does}. This activity provides an opportunity for them to reflect on the gaps in thinking \cite{milkman2009can} and revise their thoughts accordingly. For AI, this means designing appropriate mechanisms to interactively update its recommendations. For humans, the interface should possess the flexibility for them to change their WoE.
\end{itemize}

In summary, the proposed \emph{Human-AI Deliberation} architecture includes four interlinked activities and requires appropriate designs for both the decision-making interface and the AI (as shown in the dashed box in Figure \ref{fig:woe_framework} (b)). Since interface design is task-dependent, we focus on the design of \emph{Deliberative AI} in the next subsection.

\subsection{Deliberative AI: Design Considerations and Overall Structure}

Based on deliberative theories \cite{black2010methods, habermas2005concluding, thompson2008deliberative} and practices \cite{lord2013politics, steiner2005deliberative}, the Discourse Quality Index (DQI) \cite{steenbergen2003measuring} and its improved versions \cite{10.1093/oso/9780192848925.003.0006, bachtiger2009measuring} provide a comprehensive framework for assessing human deliberation. We adapt DQI to AI-assisted decision-making and derive the following design considerations (DCs):

\begin{itemize}
    \item \textbf{DC 1. Participation equality}: 
    \emph{Deliberative AI} should ensure that both parties possess equal voice \cite{chambers2005measuring} and share similar opportunities to offer opinions and reasons as well as to participate in discussions. 
    \item \textbf{DC 2. Justification rationality}: \emph{Deliberative AI} should adeptly provide rational justifications for its stances during interactions and encourage humans to do the same.      
    \item \textbf{DC 3. Constructive updates}: Rather than rigidly adhering to its initial opinions or blindly leaning towards others, \emph{Deliberative AI} should aim to facilitate compromise, reconciliation, or consensus as deliberation evolves. It should help both sides to think carefully and rationally and update their WoE in a timely manner.
    \item \textbf{DC 4. Interactivity}: \emph{Deliberative AI} should be able to understand human intentions and dynamically generate appropriate responses based on human's questions, arguments, and statements.  
    \item \textbf{DC 5. Respect and agreement}: \emph{Deliberative AI} must ensure polite discourse and respect for other participants, especially during discussions. Even if it disagrees with humans on some aspects, \emph{Deliberative AI} should show respect and understanding, creating a positive environment for continued engagement.
\end{itemize}

\begin{figure*}[htbp]
	\centering 
	\includegraphics[width=0.8\linewidth]{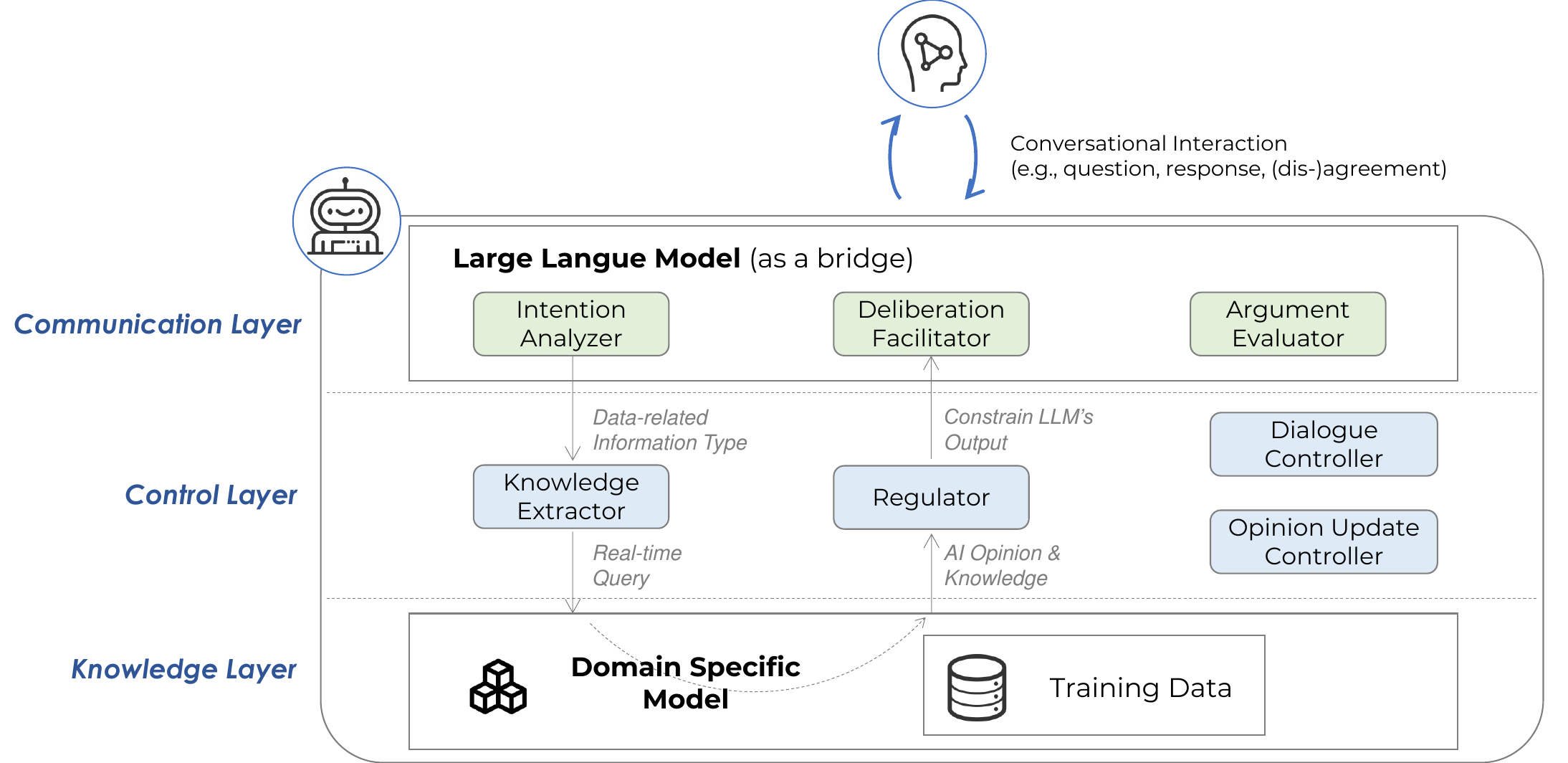}
	\caption{The architecture of \emph{Deliberative AI} which integrates a domain-specific model and a Large Language Model, enabling the AI to engage in natural communication with humans while also harnessing domain knowledge derived from the specialized model.}
	\label{fig:deliberativeAI}
        \Description{}
\end{figure*}

To fulfill these considerations, we integrate Large Language Models (LLMs) and domain-specific models (DS models) to build \emph{Deliberative AI}. DS models are responsible for the initial generation and subsequent refinement of AI's WoE. DS models' predictive power and domain knowledge offer reliable (instead of potential hallucination) information for deliberation activities. LLMs, on the other hand, bridge the interactions between humans and DS models with their conversation abilities. Overall, the architecture of \emph{Deliberative AI} (illustrated in Figure \ref{fig:deliberativeAI}) comprises three layers: \textbf{Communication} layer, \textbf{Control} layer, and \textbf{Knowledge} layer. 

\label{countrol_layer}

\begin{itemize}
    \item The \textbf{Communication} layer, empowered by LLMs, incorporates three components:
\begin{itemize}
    \item[-] \emph{Intention Analyzer} (for DC 4) understands human intent and argument evidence, facilitating cross-referencing with the Knowledge layer through the Control layer.
    \item[-] \emph{Deliberation Facilitator} (for DC 2\&5) encourages careful thinking and rational justifications while maintaining respectful deliberation.
    \item[-] \emph{Argument Evaluator} (for DC 2\&3) assesses human justification rationality, which can be used to further prompt humans’ reasoning and update AI opinions.
\end{itemize}
    \item The \textbf{Control} layer, encompassing four components, oversees:
\begin{itemize}
    \item[-] \emph{Dialogue Controller} (for DC 1) manages the deliberative discussion process (e.g., when to elicit thoughts, when to update opinions, when to move on to the next dimension, etc.)
    \item[-] \emph{Regulator} (for DC 2) guides and constrains LLM output with domain-specific model insights and training data.
    \item[-] \emph{Knowledge Extractor} (for DC 2) extracts data insights from domain-specific models and training data based on human intent analysis.
    \item[-] \emph{Opinion Update Controller} (for DC 3) adjusts AI viewpoints based on human-AI dynamics (e.g., the strength of justifications, uncertainty behind AI's opinions, etc.).
\end{itemize}
    \item The \textbf{Knowledge} layer comprises a domain-specific model and training data, providing both domain-specific knowledge and data-derived insights.
\end{itemize}

In the next section, we will provide a detailed description of how we implemented \emph{Human-AI Deliberation} and \emph{Deliberative AI} architecture in the context of a graduate admissions task.


\section{Instantiating the Architecture: Graduate Admission Prediction}

\subsection{Task, Dataset and AI Model}


We choose to use graduate admission as an illustrative task to demonstrate how to instantiate the proposed \emph{Human-AI Deliberation} architecture. In this task, participants decide on admitting or rejecting applicants to a U.S. university based on their profiles. We chose this task for two key reasons. First, this task is widely used in AI-assisted decision-making research \cite{cheng2019explaining, zhang2023deliberating, echterhoff2022ai, zhao2023evaluating}, with real-world universities employing AI algorithms for decision consistency and workload reduction \cite{waters2014grade, pangburn2019schools}. Second, graduate admission often involves deliberation among committee members \cite{zhang2023deliberating}, making it ideal for studying the effects of our proposed \emph{Human-AI Deliberation}.

The task utilizes a synthesized dataset \cite{cheng2019explaining} that simulates profiles of applicants at a U.S. public university based on publicly available aggregate statistics and distributions\footnote{Due to privacy issues, public graduate admission data sets are all synthesized. We acknowledged that it may deviate from the real-world setting.}.
The dataset comprises 100 student applications' profiles, featuring attributes considered by admission committees in actual scenarios, e.g., \emph{GRE Verbal}, \emph{GRE Quant}, \emph{GRE Writing}, \emph{GPA}, \emph{Statement of Purpose Strength}, \emph{Diversity Statement Strength}, \emph{Country}, \emph{Major}, \emph{Applicant's Undergraduate Institution Rank}, and \emph{Recommendation Letter Strength}. The dataset also includes a decision label for each case: strong reject, weak reject, weak accept, or strong accept.

To build a domain-specific model (DS-model) that can generate suggestions, we trained a multi-category linear model using a 70\% random split of the dataset as in \cite{cheng2019explaining}. We employed a linear regression model as a decision classifier, discretizing the predicted responses into one of the four decision labels. Consistent with common practices \cite{echterhoff2022ai}, we further binarized the original labels, mapping strong/weak reject to ``reject'' and strong/weak accept to ``accept'' as the ground truth for assessing AI model's and participants' prediction accuracy. The trained model achieved an 80\% accuracy on the remaining 30\% test set. The task samples used in the study were selected from the test set.



\subsection{Human-AI Thought Representation and Alignment}
In our \emph{Human-AI Deliberation} architecture, the initial step involves both humans and AI externalizing their thoughts. We employ feature contribution \cite{liao2021human, liao2020questioning} to represent their weight of evidence (WoE) along each dimension  \cite{alvarez2021human}. Feature contribution is represented by contribution scores indicating the positive or negative influence of each feature $x_i$ on the final prediction $y$. In graduate admission tasks, we treat each attribute in an applicant's profile as a dimension, and feature contribution requires humans and AI to assess the influence of each dimension on the final decision. It provides a common ground for AI and humans to express, compare, and initiate discussions.



 

For the human side, Weight of Evidence (WoE) represents humans' perceived influence of an attribute on the overall likelihood of an applicant being admitted. On the AI side, we utilized SHAP (SHapley Additive exPlanations) \cite{lundberg2017unified}, a widely-used posthoc explainable AI algorithm, to assess feature contribution/importance. SHAP values indicate both the direction and strength of a feature's impact on predictions. SHAP offers two key advantages: First, it captures feature interactions (e.g., how multiple features jointly influence the final decision), aligning with human decision-making processes. Second, its additive nature mirrors how humans combine evidence for or against options \cite{baumann2014decision}. In essence, SHAP allows feature contribution/importance to be linearly aggregate to match the model’s actual prediction no matter the AI model is linear or a complex non-linear neural network.

However, when applied to \emph{Human-AI Deliberation}, two limitations of SHAP values become evident: (1) \textbf{Interpretability of Raw SHAP Values}: Raw SHAP values (e.g., 2.15) can be difficult for non-experts to interpret directly. To address this, we converted SHAP values into probabilities by using a regression model, mapping the four-category label range to a 0-100\% scale. This conversion allows SHAP values to be directly interpreted as probabilities. (2) \textbf{Explanation of Feature Importance}: SHAP values indicate the importance of features but do not explain why a feature is important. To address this, we propose generating ``meta-explanations'' using the \emph{Knowledge Extractor} (Sec. \ref{countrol_layer}) to extract relevant evidence (e.g., data patterns) from the training data for each dimension. This approach aims to enhance the transparency of AI during the deliberation process.



Next, we describe how we implemented each component of \emph{Deliberative AI}.
\subsection{Implementation of Deliberative AI}

\subsubsection{\textbf{I. Communication Layer.}} This layer serves as a vital bridge between humans and the DS-model, facilitating effective communication by comprehending human inputs and crafting relevant responses. \ms{Note that all components in this layer are based on our designed prompts, which enable LLMs to play different roles, rather than incorporating additional predictive models.}

\emph{\textbf{I-1. Intention Analyzer}}\label{intention}. We harnessed the language capabilities of LLMs\footnote{During our experiment (conducted in August 2023), we used the GPT-3.5 model. For consistency, we will refer to it as ``LLM'' throughout this paper.} to identify human intentions and targeted dimensions in discussion. To formulate effective prompts for intention analysis, we conducted a pilot study to gather common dialogues around graduate admission decisions, including questions, arguments, critiques, and challenges. In the pilot study, we developed a preliminary version \emph{Deliberative AI} (with basic deliberative discussion capability) to carry out conversations with 30 participants from Prolific\footnote{www.prolific.co\label{prolific}}. 
Each participant expressed their opinions on various dimensions of applicant profiles and engaged in discussions with the AI. We collected 226 human deliberative statements. To extract various intentions from these statements, two authors performed qualitative coding using thematic analysis \cite{hsieh2005three}, and the results are summarized in Table \ref{intention_themes}. We then iteratively refined LLM prompts based on the collected data and built an ``Intention Analyzer'' with a 96\% accuracy in identifying themes of participant statements. Specific prompts are available in the supplementary materials.

\renewcommand{\arraystretch}{1.5}
\begin{table*}[tp]  

\centering  
\fontsize{8}{8}\selectfont  

\caption{Qualitative analysis of the sentiment/intention category of participants' statements (arguments, justifications, questions, critiques, etc.) in the deliberative discussion.}\label{table:openend}
\label{intention_themes}
\begin{tabular}{m{2.5cm}<{\centering}m{10cm}m{1.5cm}}
\toprule
Themes&Definitions and Examples&\#Participants\\
\midrule

\multirow{2}*{\shortstack{Distribution/Level of\\an attribute's values}}&\textbf{Participants evaluate how attribute values are distributed among the pool of applicants.}&\multirow{2}*{\shortstack{35 (15\%)}}\\
&``\emph{3.16 isn't a bad GPA - it's only slightly below average, sure, but it's still fairly good}'' (P2)&\\

\cline{2-3}
\multirow{3}*{\shortstack{Overall importance\\of an attribute}}&\textbf{Participants consider or challenge the overall importance of an attribute on the admission decision.}&\multirow{3}*{\shortstack{24 (10\%)}}\\

&``\emph{Diversity is extremely important to the institution as a whole so the students highly rated diversity statement would highly influence their admittance.}'' (P33)&\\

\cline{2-3}
\multirow{3}*{\shortstack{Contribution of\\an attribute}}&\textbf{Participants directly express their opinion on an attribute's contribution or challenge the contribution given by the AI but without evidence.}&\multirow{3}*{\shortstack{47 (20\%)}}\\

&``\emph{I know Applicant Undergraduate School Ranking has a significant impact on the chance of admission. But why is medium rank not good?}'' (P1)&\\

\cline{2-3}
\multirow{3}*{\shortstack{Contrastive\\evaluation}}&\textbf{Participants compare an attribute's current value with other values (often using the average) to judge an attribute's impact.}&\multirow{3}*{\shortstack{41 (18\%)}}\\

&``\emph{I am surprised you ranked the applicant's GPA on a negative scale. 3.26 is not that much lower than the 3.5 of the last applicant.}'' (P10)&\\

\cline{2-3}
\multirow{5}*{\shortstack{Holistic review\\of multiple attributes}}&\textbf{Participants evaluate how different attributes interact, taking into account the influence of certain attribute values on the strength of others.}&\multirow{5}*{\shortstack{23 (10\%)}}\\

&``\emph{The engineering major is incredibly difficult and any GPA above a 3.5 is considered successful.}'' (P3)&\\
&``\emph{I said 2\% positive influence because this individual went to a top rank school, which I assume is harder academically than some lower ranked schools.}'' (P22)&\\

\cline{2-3}
\multirow{3}*{\shortstack{Data-irrelevant\\questions/arguments}}&\textbf{Participants give data-irrelevant statements based on their heuristics, past experiences, personal beliefs, etc.}&\multirow{3}*{\shortstack{77 (34\%)}}\\



&``\emph{Statement of purpose is the only part of the application process where the applicant gets to show us who they really are in their own words - not just a score or some data value. I ranked these higher for this reason.}'' (P5)&\\

\bottomrule
\end{tabular}
\end{table*}

\emph{\textbf{I-2. Deliberation Facilitator}}. This component addresses DC2 (Justification Rationality) and DC5 (Respect and Agreement) by designing corresponding LLM prompts.
In particular, we instruct LLM to (1) Demonstrate a nuanced understanding of the human's statement; (2) analyze the specific content of the person's statement; and (3) provide a thoughtful and critical response. For detailed prompts, please refer to the supplementary materials.

\emph{\textbf{I-3. Argument Evaluator}}. The main function of this component is to assess the strength of a person's statement/argument, which later informs updates to AI opinions. Drawing from established theories in human argumentation evaluation \cite{harris1997evaluating, van2002argumentation, van2016argumentation}, we devised a comprehensive scoring mechanism with nine key items: Clarity, Relevance, Evidence, Logic, Consistency, Counterarguments, Depth, Credibility, and Alignment. These criteria are integrated into a prompt, guiding the LLM to evaluate human statements. We then average and scale the scores to obtain the overall human argument strength $S_{Human}$ (from 0 to 1; 0: weakest, 1: strongest). \ms{We conducted a pilot study to evaluate the reliability of the LLM in scoring human arguments, using data collected from our Intention Analyzer pilot study (Sec. \ref{intention} I-1). Two authors independently scored the arguments using predefined schemas, resolving disagreements to reach consensus. We then calculated Cohen’s Kappa (\(\kappa\)) to measure agreement between the LLM and human scores. The resulting \(\kappa\) value of 0.78 indicates substantial agreement, suggesting that the LLM provides reliable annotations with minimal disagreement in the context of our graduate admission task.} Additional details, including scoring schemas and prompts, can be found in the supplementary materials.

In summary, the communication layer enables general interactions with humans. To integrate it with the DS-model's predictions and knowledge, a control layer is required to connect the two.

\subsubsection{\textbf{II. Control Layer.}} This layer manages the querying and extraction of specific DS-model opinions and knowledge while controlling the entire conversation flow.

\begin{figure*}[htbp]
	\centering 
	\includegraphics[width=0.8\linewidth]{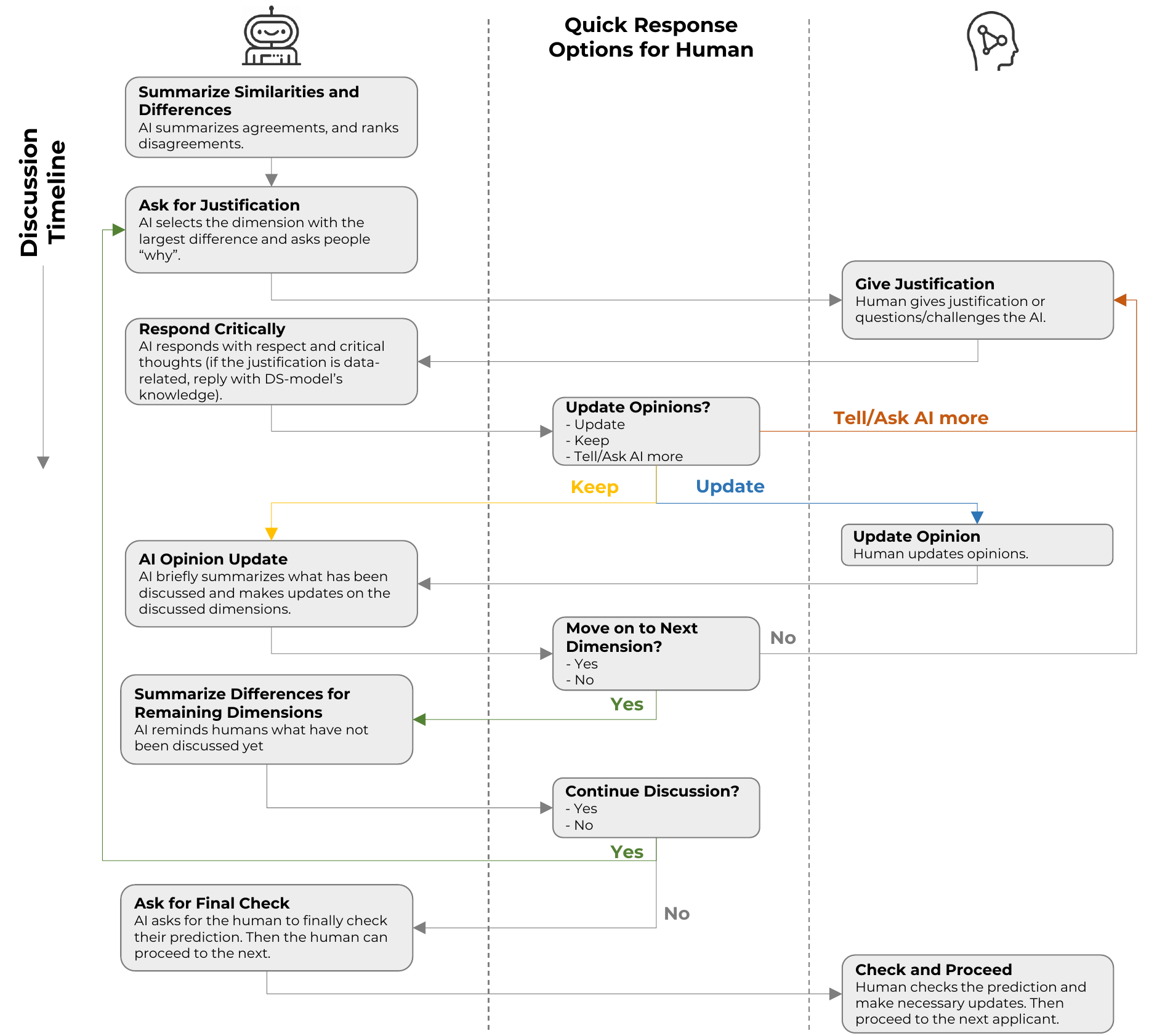}
	\caption{The conversation flow for the deliberative discussion.}
	\label{fig:conversation_flow}
        \Description{}
\end{figure*}

\emph{\textbf{II-1. Dialogue/Discussion Controller}}. 
This component serves as the control center for the discussion process, orchestrating a structured deliberation flow as shown in Figure \ref{fig:conversation_flow}. It unfolds as follows: 
[Thought Elicitation] Participants express their WoE on each dimension; AI responds with its perspectives.
[Discussion] AI highlights commonalities and discrepancies, inviting participants to provide justifications or question differing viewpoints. AI responds with critical insights. All three components of the Communication layer (\emph{Intention Analyzer}, \emph{Deliberation Facilitator}, and \emph{Argument Evaluator}) play vital roles in this phase.
After one round of discussion, AI offers input options for participants to update, maintain, or continue the discussion. AI proceeds based on participants' choices. If they wish to move to the next dimension, AI summarizes any pending dimensions, highlighting differences. Participants can choose to explore untouched dimensions, revisit previous discussions, or skip this round.
Participants have the flexibility to initiate dialogues on any dimension at any time, using quick input options or free text. They can refine their views on the decision interface independently of AI opinion updates.


\emph{\textbf{II-2. Knowledge Extractor}}. 
Based on the attributes/dimensions and intent types identified by the ``Intention Analyzer'' (see Table \ref{table:openend}), we developed a series of query functions to extract relevant data knowledge from the DS-Model. These functions help pull evidence for the LLM to generate responses in deliberative discussions appropriately. We established a mapping between the recognized intent type and the query function and called different query functions based on the recognized intent type. \ms{In practice, for a decision-making task, an existing dataset is typically available for training the AI model. This same dataset can be used to extract data patterns based on task-specific features, such as the percentile of a single feature value or the combination of multiple features. We recommend that researchers interested in applying this approach to other tasks first identify the data patterns that users are likely to find relevant, and then design the corresponding extraction functions.} Below is a brief overview of the designed functions corresponding to different human intent types. Please refer to the supplementary materials for detailed codes and examples.

\begin{itemize}
    \item \emph{Distribution/Level of an attribute's value}:
    \begin{itemize}
        \item Function \texttt{get\_distribution(\textcolor{gray}{attr\_val})} calculates attribute value percentiles within the applicant pool, along with contextual comparisons (with minimum, maximum, quartiles, mean, and median).
    \end{itemize}
    
    \item \emph{Overall attribute importance}:
    \begin{itemize}
        \item Function \texttt{get\_global\_feature\_importance(\textcolor{gray}{attr})} returns global importance.
        \item Function \texttt{get\_correlation(\textcolor{gray}{attr})} provides Pearson correlation between the selected attribute and the admission chance.
        \item Function \texttt{get\_influence\_on\_admission\_chance(\textcolor{gray}{attr})} calculates admission chance changes for varying attribute values.
    \end{itemize}

    \item \emph{Contribution of an attribute}:
    \begin{itemize}
        \item Function \texttt{get\_current\_value\_influence(\textcolor{gray}{attr})} calculates admission chance differences when the value of a selected attribute is randomized.
    \end{itemize}

    \item \emph{Contrastive Evaluation}:
    \begin{itemize}
        \item Function \texttt{get\_contrastive(\textcolor{gray}{attr, contrast})} computes admission chance differences compared to a contrastive value.
    \end{itemize}

    \item \emph{Holistic review of multiple attributes}:
    \begin{itemize}
        \item Function \texttt{get\_holistic\_analysis(\textcolor{gray}{attr, fix\_attr})} evaluates the impacts of attribute(s) in specific scenarios, e.g., the impact of the GPA percentile in [top-ranked universities].
    \end{itemize}
\end{itemize}


\emph{\textbf{II-3. Regulator}}. The primary objective of this component is to harness the expertise of the DS-Model to regulate the responses generated by LLM. This approach makes certain that LLM's responses always align with the DS-Model's knowledge and decisions.
To achieve this goal, we created consistency-ensuring prompts based on three key elements: (1) the findings extracted by the \emph{Knowledge Extractor}, (2) the overarching decisions made by the DS-Model, and (3) the DS-Model's viewpoint on the current attribute under discussion. 

\emph{\textbf{II-4. Opinion Update Controller}}: We updated the AI's opinions by taking into consideration: (1) the current opinions of both the human ($O_{Human}$) and the AI ($O_{AI}$) on the discussed attribute, (2) the strength of the human's argument ($S_{Human}$, see \emph{Argument Evaluator}), and (3) the AI's uncertainty ($U_{AI}$) measured and calibrated via Uncertainty Quantification 360 toolbox \cite{ghosh2021uncertainty} (the uncertainty ranges from 0 to 1: the closer to 1, the more uncertain AI's prediction is). 
We propose the following formula to update the AI's opinions ($\hat{O}_{AI}$) on an attribute based on these factors, inspired by result aggregation in crowd intelligence \cite{lyon2013wisdom, galton1907vox}. \ms{It is important to clarify that the term ``update'' here does not refer to modifications to the domain-specific model itself, such as retraining or fine-tuning. Instead, it pertains to the adjustment of the AI’s expressed viewpoints regarding the current task case. Notably, LLM is not directly involved in updating the domain-specific model's viewpoints. Rather, LLM generates a strength score for the human's arguments, with the update being performed using Eq. \ref{equation1}}.

\begin{small}
\begin{equation}
    \hat{O}_{AI} = \frac{1 - U_{AI}}{1 - U_{AI} + S_{Human}} \cdot O_{AI} + \frac{S_{Human}}{1 - U_{AI} + S_{Human}} \cdot O_{Human},
\label{equation1}
\end{equation}
\end{small}


\subsubsection{\textbf{III. Knowledge Layer}} 

This layer comprises the DS-Model and the training dataset. The DS-Model provides overall predictions and opinions on each dimension, while the training dataset supplies essential information (e.g., data distributions and patterns) for the \emph{Knowledge Extractor} to perform real-time calculations and queries.

Overall, in this architecture, LLM is used for language understanding and generation. The opinions and evidence used by LLM are retrieved in real time from the DS-Model and training data through our logic code (like retrieval augmented generation \cite{lewis2020retrieval}). In this way, the LLM is used in a controllable and responsible manner, minimizing the potential hallucination. We provide an example of how data is processed in Deliberative AI in the Appendix.

\subsection{Interface Design}

The interface for the graduate admission task is structured into three main regions:
\begin{figure*}[htbp]
	\centering 
	\includegraphics[width=\linewidth]{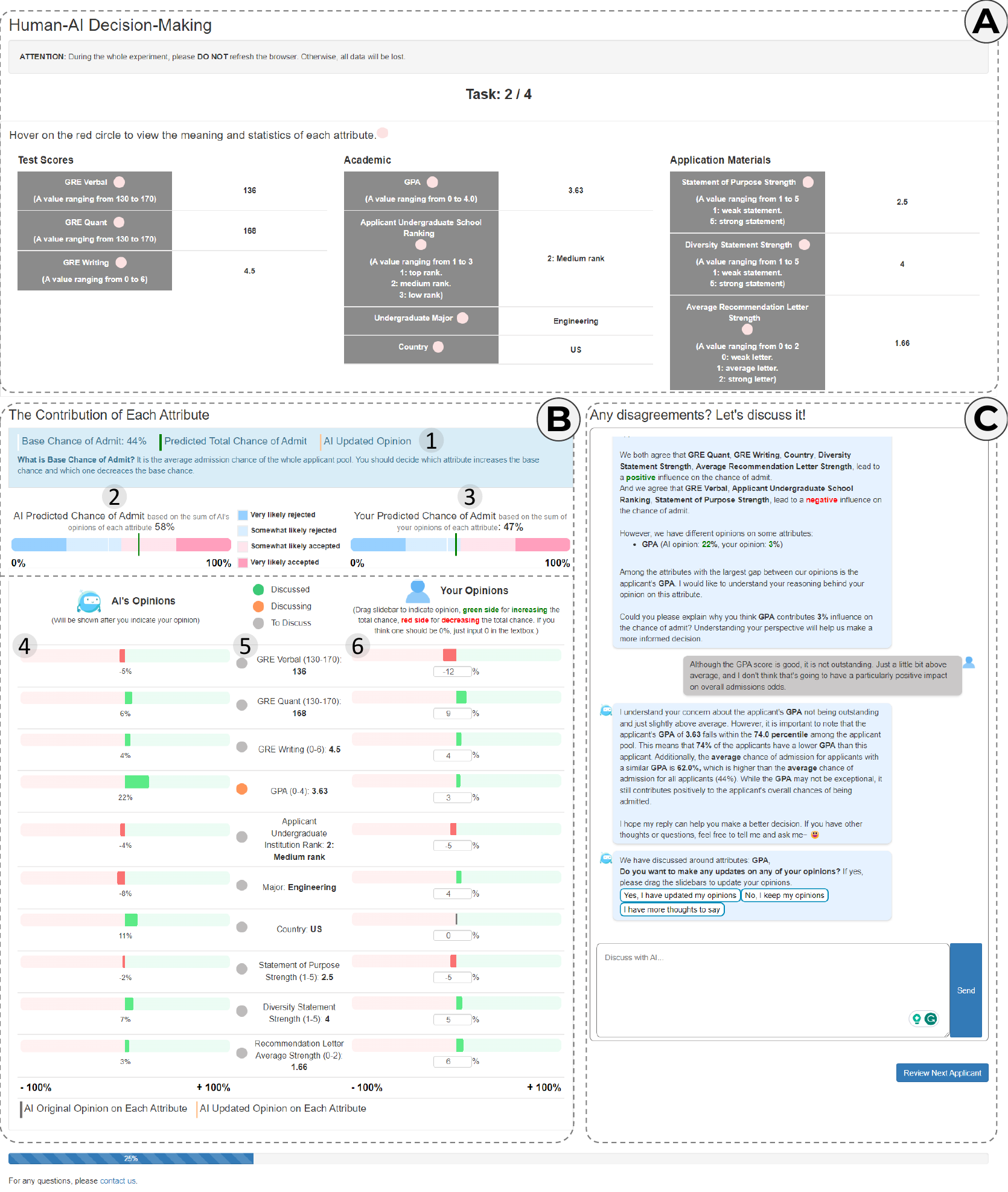}
	\caption{The interface of \emph{Deliberative AI}. The interface contains three parts. The top part (A) is the applicant's profile. The bottom left part (B) is the region for humans and AI to indicate (and update) their opinions. The bottom right part (C) is the discussion region where humans and AI can discuss conflicting opinions. (All the dashed lines are only for illustration)}
	\label{fig:interface}
        \Description{}
\end{figure*}

\begin{itemize}
    \item \textbf{Profile Region} (Figure \ref{fig:interface} (A)) displays the applicant's profile, providing a table with the current value and possible range of each attribute. Users can access attribute definitions and basic data distribution statistics (minimum, maximum, average, and median values) by hovering over pink circular markers.
    \item \textbf{Opinion and Prediction Region} (Figure \ref{fig:interface} (B)) is dedicated to thought elicitation by both users and the AI.
\begin{itemize}
    \item[-] The upper part displays aggregate predictions from both humans and AI. This includes a legend (Figure \ref{fig:interface}-1) and two slide bars (Figure \ref{fig:interface}-2 and -3) representing AI's and the user's overall predictions, respectively. Each slide bar shows three line indicators: a white line representing the average admission probability of all applicants, a green line showing the initial predictions made by humans/AI, and a yellow line denoting the updated prediction by humans/AI (only shown after an update is made).
    \item[-] The bottom part of this region enables both humans and AI to express opinions on each decision dimension (i.e., applicant attribute). A simplified profile in the middle (Figure \ref{fig:interface}-5) reduces attention shifts. Status indicators show if a dimension has been discussed (green), is being discussed (orange), or is yet to be discussed (gray). Separate “concrete opinion” sliders (Figure \ref{fig:interface}-4 and Figure \ref{fig:interface}-6) allow AI and humans to share dimension-level opinions.
    \item[-] Each dimension's slide bar starts in a central position (0\% contribution).
    Users can drag the slider any time to the right to increase the weight on an attribute toward a positive ``admit'' decision or to the left to reduce its contribution. Alternatively, users can directly input contribution values in a box below the slider.  
\end{itemize}
Slide bars in Figure \ref{fig:interface}-2 and Figure \ref{fig:interface}-4 are interconnected, so as those in Figure \ref{fig:interface}-3 and Figure \ref{fig:interface}-6. Values within the ``overall prediction bar'' reflect the cumulative values from the ``concrete opinion bars.'' Any changes in the dimension-level bars immediately update the overall prediction. 
AI's and the user's opinions are displayed side by side for easy comparison. 
Note that at the beginning of each case, users have to complete their opinion inputs and click the [Submit Opinion] button to see AI's initial (overall and concrete) suggestions.

    \item \textbf{Discussion Region} (Figure \ref{fig:interface} (C)) is where all deliberative dialogues take place. Users can type out their opinion arguments, questions, disagreements with AI, responses to AI queries, and more. Importantly, changes made in the Opinion Region are seamlessly integrated by AI and reflected in ongoing discussions, and conversely, any viewpoint changes mentioned in the dialogue are instantly updated in the Opinion Region.
\end{itemize}


\section{Exploratory User Study}
To gain an initial understanding of the impact of the \emph{Human-AI Deliberation}, we conducted a mixed-methods study within the context of graduate admissions. This study is termed \emph{exploratory} because our objective was not to assess the effectiveness of \emph{Human-AI Deliberation} specifically for graduate admissions. Rather, we used graduate admissions as an illustrative task to explore the potential effects of \emph{Human-AI Deliberation} on AI-assisted decision-making.


\subsection{Task Setup}
We used the graduate admission task as our testbed. To ensure manageable study durations and prevent participant fatigue, we selected five task cases based on the length of the pilot study. These cases included one for the tutorial and four for the main tasks. To investigate human-AI deliberation, we selected cases from a pilot study where human and AI opinions conflicted, often involving ambiguous applicant profiles near the admission borderline. As a result, predicting the admission outcome for these cases was challenging for both humans and AI, leading to performance levels between 50\% and 60\% for both. However, this does not imply that the AI model used in our study is of unrealistically low performance; rather, we focused on ambiguous task cases that require conflict resolution.


\subsection{Conditions}
We compared the proposed \emph{Deliberative AI} with the traditional explainable AI assistant (\emph{XAI}) and \emph{Human Alone}.
\begin{itemize}
    \item \textbf{Deliberative AI (DAI)}: Participants share their thoughts on various dimensions before viewing AI recommendations. We present AI's ``thoughts'' on each dimension afterward. After comparing conflicting viewpoints, we offer a dialogue interface for participants and AI to discuss any of the perspectives, as shown in Figure \ref{fig:interface}.
    \item \textbf{Explainable AI (XAI)}: After individuals provide their predictions, they receive AI recommendations (along with feature contribution-based explanations) and then make their final judgments (see Figure \ref{fig:baselineinterface} in Appendix).
    \item \textbf{Human Alone}: Participants need to make predictions independently without any AI assistance.
\end{itemize}

\subsection{Procedure}

With our institutional IRB approval, we conducted a between-subjects study. After obtaining consent, we had participants complete a background questionnaire to gather demographic data and assess their AI expertise. \ms{We then introduced the study, explained the task, workflow, and AI’s functions, including its ability to update opinions, without delving into the specifics of the adjustment mechanism. This approach reflects real-world scenarios, where non-technical users focus more on functionality than technical details.} Participants then engaged in an interactive tutorial, practiced with one example task, and received distribution and summary statistics for each attribute of the applicant's profile. After the tutorial, we asked qualification questions to check participants' understanding of the task, allowing only those who answered all questions correctly to proceed to the main task. In the main task, participants worked on four graduate admission task cases, which were presented randomly. Finally, we collected participants' perceptions, experiences, and feedback on the AI system and the discussion process in the exit survey.

\subsection{Participants}

We first conducted a power analysis to determine the required sample size for using G*Power \cite{faul2009statistical} with a default effect size $f$=0.25 (indicating a moderate effect), a significance threshold $\alpha$=0.05, and a statistical power $1-\beta$=0.8. This resulted in a required total sample size of 159 participants for the three conditions. After obtaining institutional IRB approval, we recruited a total of 174 participants from Prolific\textsuperscript{\ref{prolific}}.
To ensure high-quality responses, participants had to meet specific criteria: (1) residing in the United States; (2) having been admitted to a US graduate program before (as the task involved predicting graduate admission in a US university); (3) having at least a 99\% approval rate with at least 1000 previous submissions; (4) using English as their first language; and (5) using a desktop computer for the experiment. After filtering based on attention-check questions, we obtained 153 valid responses (\emph{Deliberative AI}: 48, \emph{XAI}: 51, \emph{Human Alone}: 54).
Among the final participants, 84 self-identified as male, 67 as female, and 2 as others. The age distribution was as follows: 23 participants aged 24–29, 42 aged 30–39, 33 aged 40–49, 30 aged 50–59, and 25 aged over 59. \ms{Regarding education, 125 participants held an MA/MSc/MPhil degree, and 28 held a Doctorate (PhD or equivalent)}. Participants also had diverse levels of AI knowledge: 9 reported having no knowledge, 86 were familiar with basic AI concepts, 50 had experience using AI algorithms, and 8 identified as AI experts. Participants in the \emph{Deliberative AI} condition received bonuses based on the actual study length. To motivate high-quality work, participants received a \$0.50 bonus if their overall accuracy exceeded 75\%. On average, participants earned about \$12 per hour.

\subsection{Measurement}
To answer the aforementioned research questions, we comprehensively measured the effects of human-AI deliberation across four dimensions: \emph{task performance}, \emph{reliance}, \emph{perceptions of AI}, and \emph{user experience}.

\begin{itemize} \item \textbf{Task Performance}. We evaluated decision-making accuracy using \emph{Decision Accuracy} metrics \cite{zhang2020effect, bansal2021does}.
\item \textbf{Reliance}. Participants' reliance on AI suggestions was assessed through the \emph{Agreement Fraction} \cite{zhang2020effect, he2023knowing} and \emph{Switch Fraction} \cite{zhang2020effect, he2023knowing}. Additionally, the appropriateness of reliance was measured using the \emph{Over-reliance Ratio} \cite{wang2021explanations, visser2014design, parasuraman1997humans} and \emph{Under-reliance Ratio} \cite{wang2021explanations, visser2014design, parasuraman1997humans}.  

\item \textbf{Perceptions of AI}. Participants' perceptions of AI were measured using 7-point Likert scales for \emph{Helpfulness} \cite{laugwitz2008construction, cai2019human, buccinca2020proxy}, \emph{Trustworthiness} \cite{ghai2021explainable, buccinca2021trust}, and \emph{Understanding} \cite{wang2021explanations} (1: Strongly disagree; 7: Strongly agree).  

\item \textbf{User Experience}. Participants' \emph{Decision Confidence} was evaluated using established measures \cite{miller2015meta}. Given that deliberation requires additional effort, we also assessed \emph{Mental Demand} \cite{hart2006nasa, lai2022human, ghai2021explainable, buccinca2021trust}, \emph{Effort} \cite{hart2006nasa}, \emph{Complexity} \cite{buccinca2021trust}, and \emph{Satisfaction} \cite{buccinca2021trust, ghai2021explainable}, all measured on 7-point Likert scales.
\end{itemize}

To gain a deeper understanding of participants' perceptions of both \emph{Deliberative AI} and the deliberative decision-making process, we also gathered open-ended feedback in the exit survey. These questions explored participants' perceptions of the usefulness of deliberating with AI, the AI's updates, and their expectations for system improvements. A detailed overview of the metrics and questions is provided in Table \ref{tab:measurement} in the Appendix.

\subsection{Analysis Methods}
We conducted mixed-methods analyses to examine the aforementioned metrics. For quantitative analysis, we first performed normality tests (Shapiro-Wilk) and found that the data did not fit the normality assumption. Therefore we ran the non-parameter tests. Specifically, to compare \emph{Deliberative AI} and \emph{XAI} (such as humans' reliance on AI, and their perceptions of AI), we run Mann-Whitney U test. To compare all three conditions (such as task performance, and user experience), we employed Kruskal–Wallis tests with Bonferroni post-hoc correction and we reported adjusted p-values.

For qualitative analysis, two authors independently coded participants' open-ended feedback and conversation logs using an inductive thematic analysis approach \cite{hsieh2005three}. The final themes emerged through discussions and harmonization over several iterations. We also identified representative examples from the source texts for demonstration in this paper.
\section{Results}

In this section, we report our exploratory findings regarding the three research questions: (\textbf{RQ1}) how \emph{Human-AI Deliberation} affects task performance and human reliance (and reliance appropriateness) on AI, (\textbf{RQ2}) how \emph{Human-AI Deliberation} affects human perceptions and task experience, and (\textbf{RQ3}) how humans perceive the effectiveness of \emph{Human-AI Deliberation} and what should be improved.



\subsection{RQ1: How will \emph{Human-AI Deliberation} affect task performance and humans' reliance (and its appropriateness) on AI suggestions?}    
\subsubsection{Decision Accuracy} 
\begin{figure*}[htbp]
	\centering 
	\includegraphics[height=3.5cm]{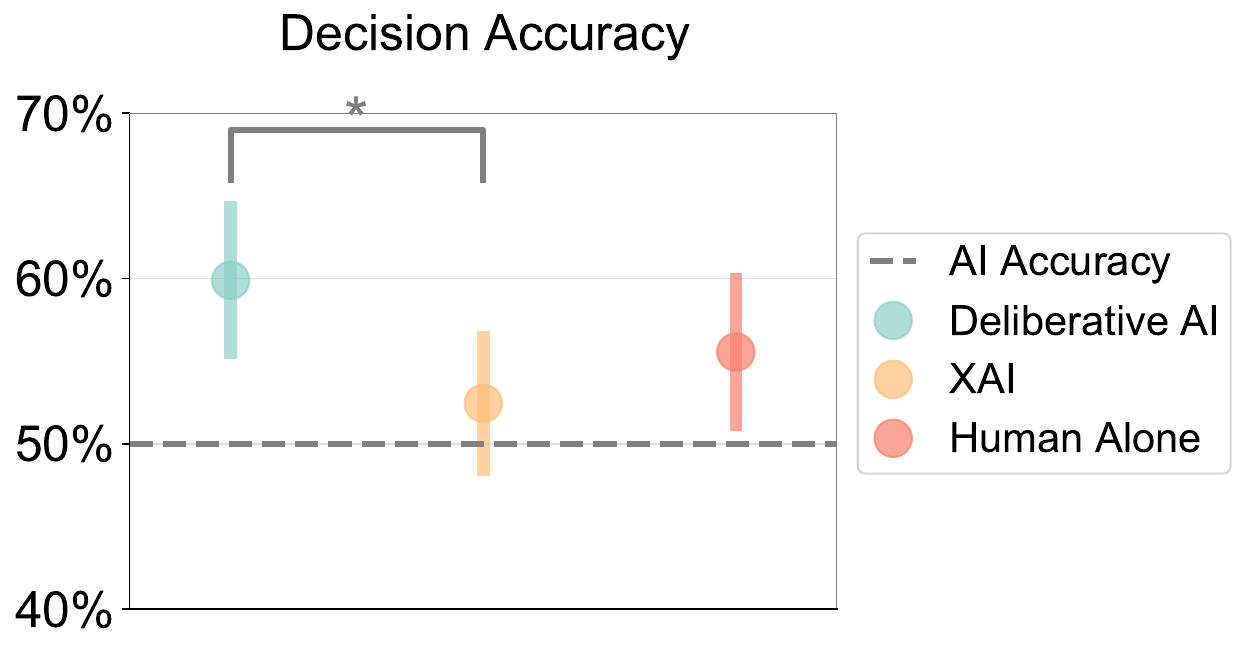}
	\caption{Task performance in different conditions. It is important to note that we intentionally selected ambiguous task cases that are prone to conflicts between humans and AI and are challenging for both. As a result, the accuracy of both humans and AI individually is relatively low. The error bars represent 95\% confidence interval. (*: $p<$0.05)}
	\label{fig:performance}   
        \Description{}
\end{figure*}

As shown in Figure \ref{fig:performance}, participants in the \emph{Deliberative AI} condition demonstrated significantly higher decision accuracy ($M$=0.598, $SD$=0.169) compared to those in the \emph{XAI} condition ($M$=0.524, $SD$=0.16, $p<$0.05). This finding suggests that in scenarios where tasks are challenging for both humans and AI—where the individual performance of humans and AI is relatively low—traditional Explainable AI (XAI) may not improve performance and could even have negative effects. In contrast, \emph{Human-AI Deliberation} shows potential to yield positive outcomes, even when the performance of the underlying AI models in these difficult task cases is suboptimal.



\begin{figure*}[h]
\centering
\subfigure[]{
\begin{minipage}[t]{0.43\linewidth}
\centering
\includegraphics[height=3.5cm]{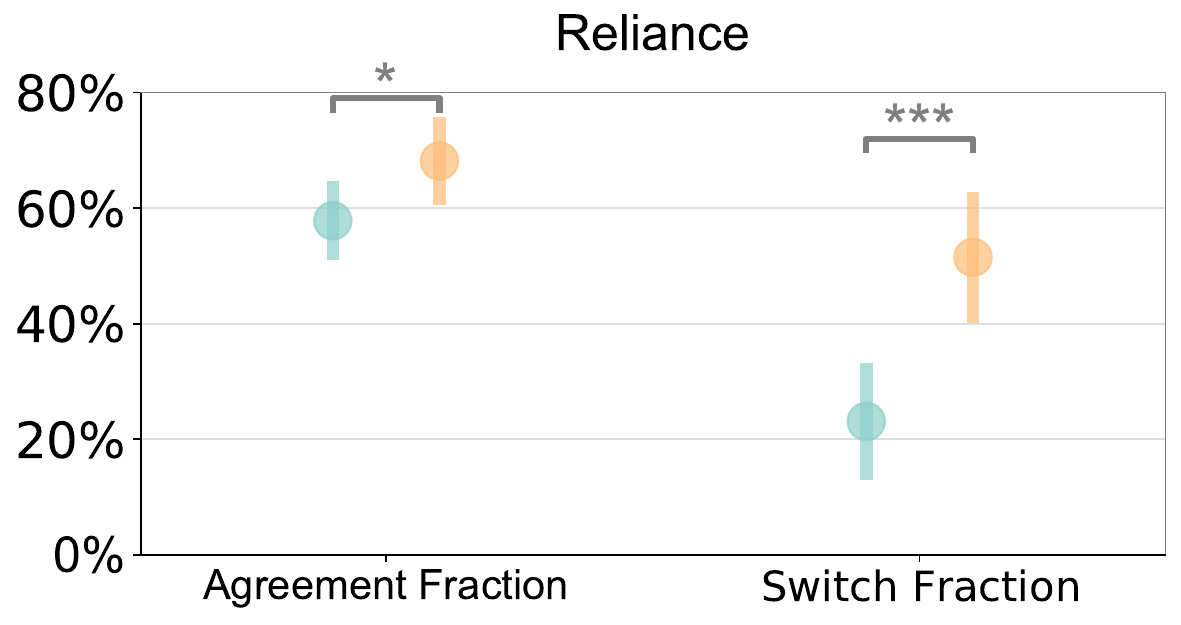}  
\end{minipage}%
}%
\subfigure[]{
\begin{minipage}[t]{0.57\linewidth}
\centering
\includegraphics[height=3.5cm]{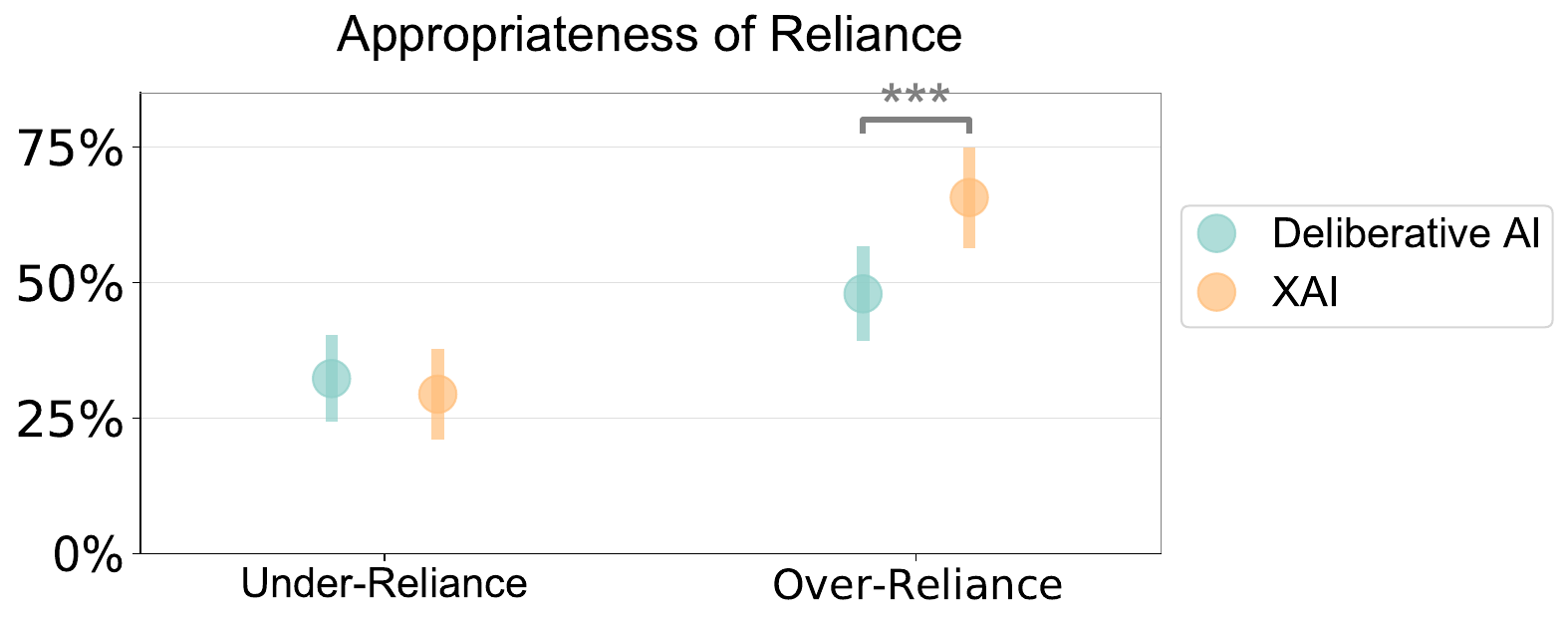}  
\end{minipage}%
}%
\centering
	\caption{Participants' reliance and the appropriateness of their reliance. (A) Participants' reliance on AI's suggestions was measured by agreement fraction and switch fraction. (B) The appropriateness of participants' reliance on AI's suggestions, including under-reliance (the ratio where participants did not use a correct AI suggestion) and over-reliance (the ratio where participants used an incorrect AI suggestion). The error bars represent 95\% confidence interval. (*: $p<$0.05, ***: $p<$0.001)}
	\label{fig:reliance}  
\end{figure*}

\subsubsection{Reliance}
\label{sec:reliance}
We measured participants' objective reliance by agreement fraction and switch fraction. As shown in Figure \ref{fig:reliance} (a), participants agreed significantly less with AI's suggestions in \emph{Deliberative AI} ($M$=0.57, $SD$=0.24) than in \emph{XAI} ($M$=0.68, $SD$=0.27, $p<$0.05), and switched significantly less to AI's predictions in \emph{Deliberative AI} ($M$=0.23, $SD$=0.35) than in \emph{XAI} ($M$=0.51, $SD$=0.41, $p<$0.001). Combined with participants' open-ended feedback (Sec. \ref{open-end}), this may be because people invest more in independent thinking in the process of deliberation with AI and realize the problematic aspects of AI's perspective.

\subsubsection{Appropriateness of Reliance} We further measured the appropriateness of participants' reliance on AI's suggestion by under-reliance and over-reliance (Figure \ref{fig:reliance} (b)). Results show that there is no significant difference between \emph{Deliberative AI} ($M$=0.32, $SD$=0.28) and \emph{XAI} ($M$=0.29, $SD$=0.30) in terms of under-reliance. While significantly less over-reliance was observed in \emph{Deliberative AI} ($M$=0.47, $SD$=0.31) than in \emph{XAI} ($M$=0.65, $SD$=0.33, $p<$0.001), which means that participants had more appropriate reliance on AI when collaborating with our proposed \emph{Deliberative AI}. \ms{This result aligns with existing research on cognitive forcing functions \cite{buccinca2021trust}, where high cognitive effort reduces over-reliance on AI. However, unlike in \cite{buccinca2021trust}, where cognitive effort significantly increased under-reliance (i.e., humans under high cognitive effort may blindly ignore AI's suggestions), our findings reveal no such adverse effect. This indicates that our tool mitigates over-reliance not merely by increasing cognitive effort but by fostering meaningful deliberation.}




\subsection{RQ2: How will \emph{Human-AI Deliberation} affect humans' perceptions of the AI and user experience?}
We measured the effects of different AI conditions on participants' perceptions and user experience via a 7-point Likert scale (1: strongly disagree, 7: strongly agree).

\begin{figure*}[h]
\centering
\subfigure[]{
\begin{minipage}[t]{0.35\linewidth}
\centering
\includegraphics[width=0.95\linewidth]{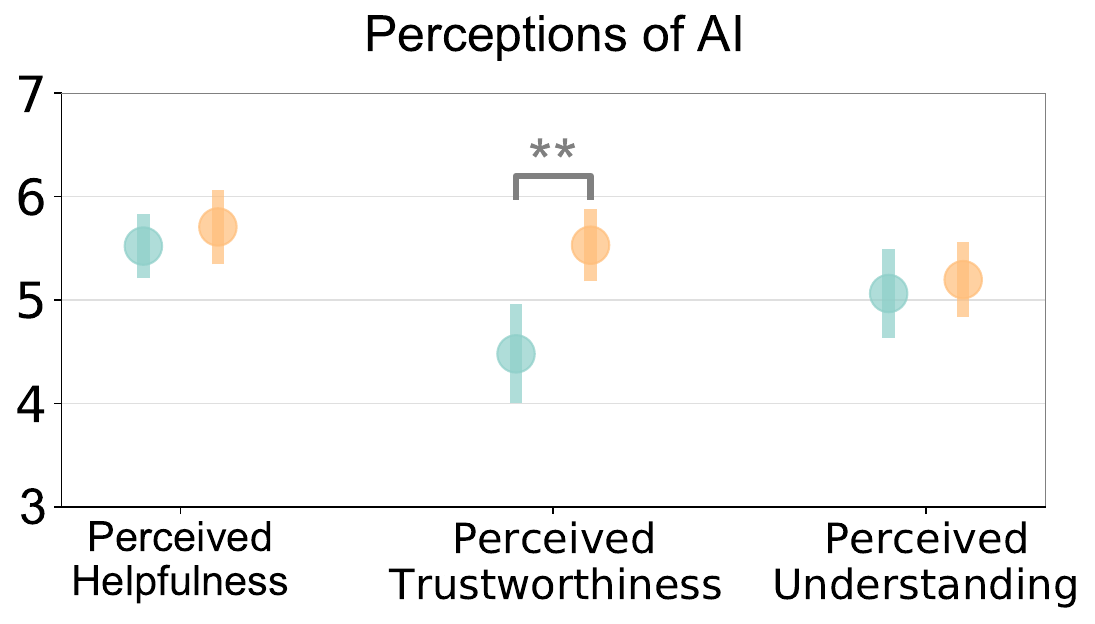}
\end{minipage}%
}%
\subfigure[]{
\begin{minipage}[t]{0.65\linewidth}
\centering
\includegraphics[width=\linewidth]{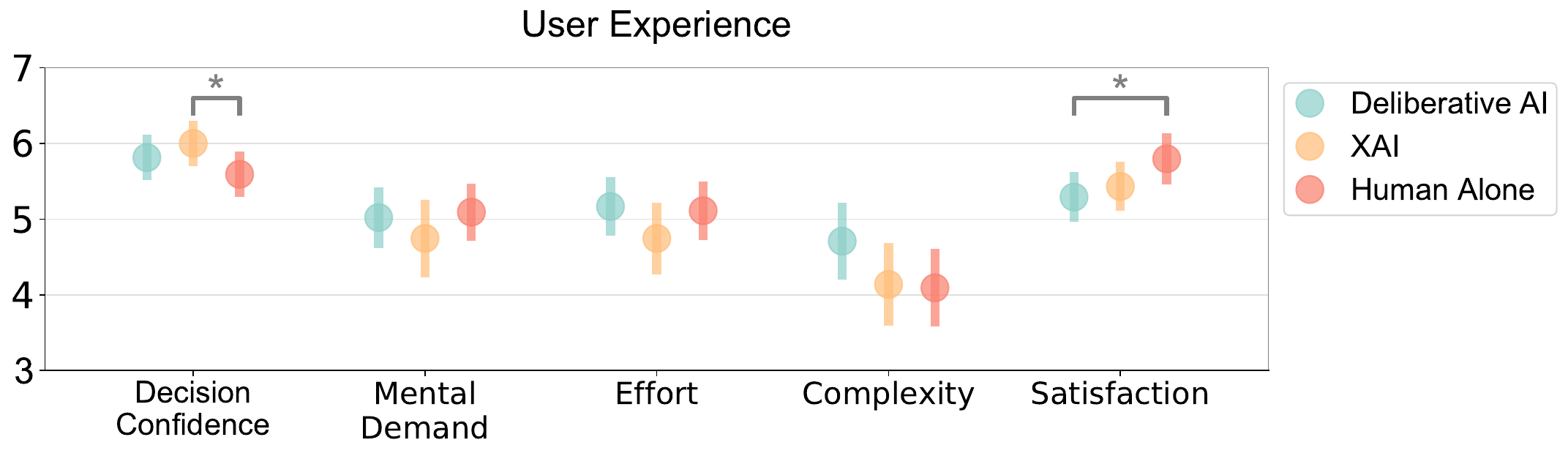}
\end{minipage}%
}%
\centering
	\caption{Participants' perceptions and task experience. (A) Participants' perceptions of different AI assistant. (B) Effects on user experience. The error bars represent 95\% confidence interval. (*: $p<$0.05, **: $p<$0.01)}
	\label{fig:perception_ux}  
\end{figure*}


\subsubsection{Perceptions of AI}
Figure \ref{fig:perception_ux} (a) shows participants' perceptions of the AI model. There were no significant differences in \emph{perceived helpfulness} and \emph{understanding} between \emph{Deliberative AI} and \emph{XAI}. However, participants reported significantly less trust in \emph{Deliberative AI} ($M$=4.47, $SD$=1.68) compared to \emph{XAI} ($M$=5.52, $SD$=1.27, $p<$0.01), aligning with their reliance behaviors (Sec. \ref{sec:reliance}). This difference may be attributed to participants identifying more AI flaws through deliberation than by solely observing AI's explanations, supported by conversation logs analysis (see Sec. \ref{open-end} for details).

\subsubsection{User experience}
\label{sec_ux}
First, we want to see participants' decision confidence. As indicated in Figure \ref{fig:perception_ux} (b), participants in \emph{XAI} reported significantly higher confidence ($M$=6, $SD$=1.10) in their predictions than those in \emph{Human Alone} ($M$=5.59, $SD$=1.12, $p<$0.05). However, from Figure \ref{fig:performance} (a) we found that the final accuracy of participants in \emph{XAI} is even lower than those in \emph{Human Alone}. This indicates that the traditional \emph{XAI} might lead to humans' \emph{\textbf{illusionary confidence}}, which could prevent humans from making optimal decisions. 

Given that the \emph{Deliberative AI} requires participants to externalize thoughts at a dimension level and engage in deliberative discussions on conflicting opinions, it's crucial to explore how these activities influence the user experience. Results showed no significant difference among the three conditions concerning \emph{Mental Demand}, \emph{Effort}, and \emph{Perceived System Complexity}. However, we find participants reported significantly lower \emph{Satisfaction} in \emph{Deliberative AI} than in \emph{Human Alone}. This decrease in user experience may be due to the AI exposing more conflicts for humans to resolve. As P3 noted, ``\emph{It’s annoying because I have to try to find evidence to prove that my point of view is correct.}'' This result suggests that there is a trade-off between encouraging users' deliberative thinking and optimizing their user experience, in line with findings in previous studies \cite{buccinca2021trust}. Future work should focus on finding a balance between the benefits of deliberation and maintaining a positive user experience.

\subsection{RQ3: How will humans perceive the effectiveness of the proposed \emph{Human-AI Deliberation} and what can be improved for future \emph{Human-AI Deliberation} design?}
\label{open-end}
\begin{figure*}[htbp]
	\centering 
	\includegraphics[width=\linewidth]{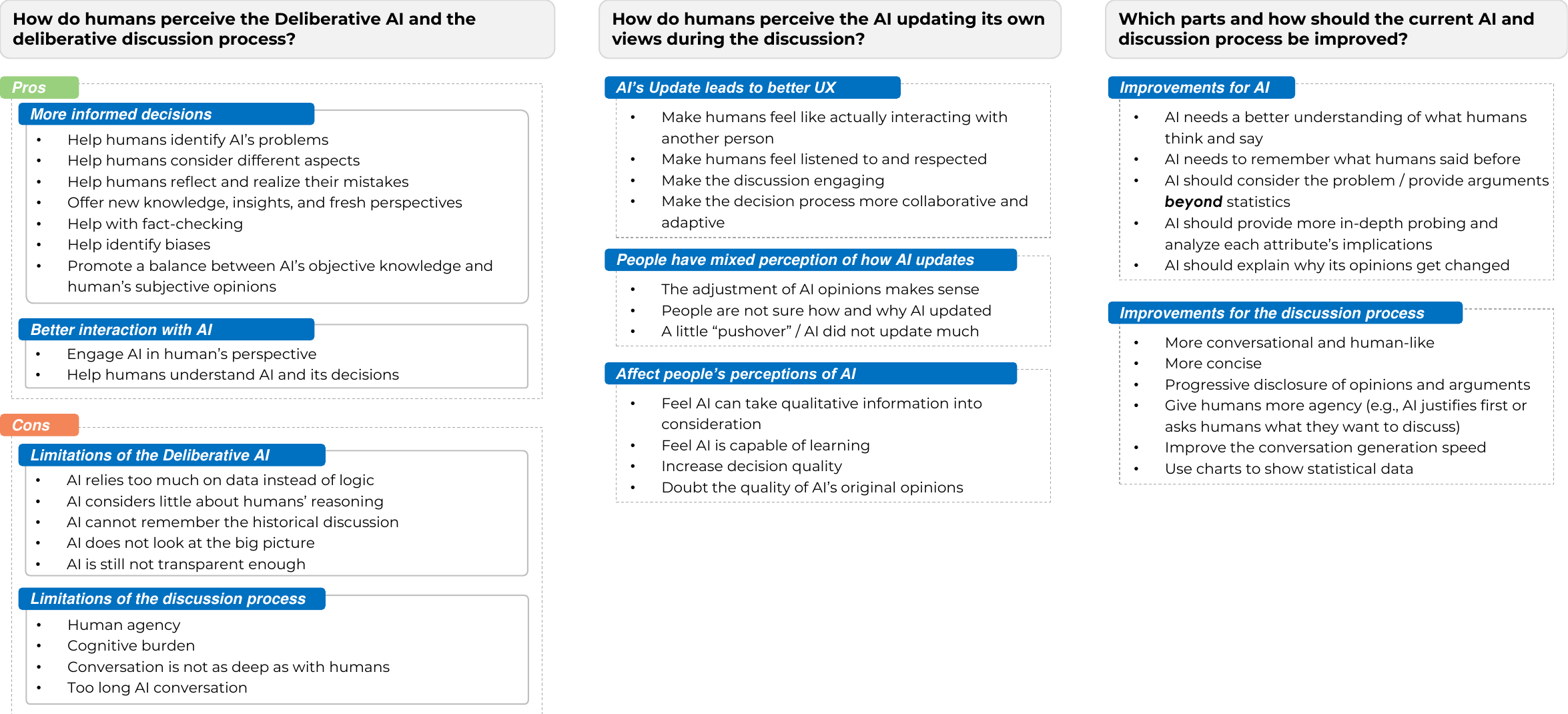}
	\caption{The main results of our thematic analysis of the open-ended questions.}
	\label{fig:qualitative}
        \Description{}
\end{figure*}

In addition to quantitative measures, we aimed to gain a deeper understanding of participants' perceptions of the helpfulness of the proposed \emph{Human-AI Deliberation} and the \emph{Deliberative AI} feature, particularly the opinion-updating aspect. We also sought insights to inform future design improvements for \emph{Human-AI Deliberation}. To achieve this, we analyzed participants' open-ended feedback, supported by their conversation logs. Our key findings are summarized in Figure \ref{fig:qualitative}. Below, we present key insights, with themes highlighted in bold.

\subsubsection{Perceptions of Deliberative AI and the Discussion Process}
Participants offered both positive and critical feedback on \emph{Deliberative AI} and the deliberation process. Of the 48 participants who experienced with \emph{Deliberative AI} condition, 43 felt that \textbf{deliberation helped them make more informed decisions}. Specifically, AI-assisted discussions enabled participants to “\textbf{identify AI's limitations}” (21/48), “\textbf{consider different perspectives}” (10/48), and “\textbf{reflect on and correct their own mistakes}” (15/48). For example, P1 (Male, 32) pointed out the AI’s overreliance on GPA scores while underestimating the role of recommendation letters: \emph{"The AI relies too much on GPA scores but undervalues recommendation letters. It didn’t provide convincing justifications, so I couldn’t rely on its opinion for these factors."} P19 (Male, 42) noted how the deliberation process encouraged self-reflection: \emph{"The AI made me question what I believed to be sufficient reasoning, especially given the data."} These reflections are consistent with our analysis of participants' conversation data with \emph{Deliberative AI}, where participants expressed doubts to the AI in 32\% of dialogues, acknowledged its arguments in 17\% of dialogues, and engaged in self-reflection and correction in 15\% of dialogues. 

Moreover, 18 participants found that \textbf{discussing with the AI introduced “new knowledge, insights, and perspectives”}. For instance, P5 (Male, 45) commented: \emph{"The AI provided information I didn’t know, like percentiles and how similar stats influenced past decisions, which I found extremely helpful."}

Nine participants said that \textbf{deliberation helped them recognize biases}. For example P35 (Female, 29) mentioned: \emph{"It made me realize the AI had inherent biases, which prompted me to pause and reflect."} This aligns with findings from the participants’ conversation data with AI, where many participants (15/48) identified biases in the AI’s reasoning. For example, the AI exhibited bias by assigning more importance to U.S. applicants while downplaying those from Asia. P3 (Male, 42) questioned: \emph{"Why does their country matter? Penalizing someone for their nationality seems biased."}

Furthermore, 15 participants appreciated that \textbf{deliberation balanced the AI’s objective data with human subjective judgment}.” For instance, P2 (Male, 42) said: \emph{"It provides an opportunity to consider multiple perspectives and statistical data, resulting in more balanced decision-making."}

Despite the benefits, participants also identified limitations in the deliberative process. Five participants felt that their \textbf{sense of agency was reduced as the AI prompted them to think, rather than passively waiting for their input}. Four participants found the \textbf{discussion is mentally demanding} and criticized the AI’s verbose responses, reflecting the results on user experience (see Sec. \ref{sec_ux}).

Regarding the limitations of \emph{Deliberative AI}, two main concerns emerged: that “\textbf{Deliberative AI relied too heavily on data over logic}” (15 participants) and “\textbf{failed to consider human reasoning}” (10 participants). This feedback aligns with our analysis, where participants integrated personal experiences and logic into their decisions, while the AI lacked this depth. For example, participants considered factors such as: \emph{"Business programs are highly competitive, requiring a higher GPA."} (P17, Female, 60) and \emph{"A strong SOP reflects a deep understanding of the project and strong commitment."} (P32, Male, 28)

Additional concerns included the AI’s insufficient ability to faithfully remember previous discussions (five participants), failure to consider the broader context (three participants), and lack of transparency (three participants).


\subsubsection{Perceptions of \emph{Deliberative AI}'s Updates}
Participants generally appreciated the dynamic updating feature. 25 out of 48 participants felt that \textbf{the AI updates improved the user experience}. They described the experience as more interactive, comparing it to “\emph{interacting with a real person}” (6 participants), where their opinions felt “\emph{heard and respected}” (19 participants). Moreover, these updates made the discussion “\emph{more engaging}” (8 participants) and the decision-making process “\emph{more collaborative}” (5 participants).

However, perceptions of AI’s updates varied. While 11 participants felt the AI adjusted its views too frequently (e.g., “\emph{the AI updated too much}”), 13 participants expressed uncertainty about “\emph{why and how the AI was updating}.” For example, P2 (Male, 42) noted: \emph{"I'm not sure if the AI really changed its opinion based on what I said or if it was just programmed to do so in response to my input."} Interestingly, \textbf{some participants thought the AI was too quick to change its stance, while others felt it wasn’t flexible enough}. These updates also affected participants' perceptions of the AI itself. Eight participants felt the AI could “\emph{incorporate qualitative information},” and five believed the AI was ``\emph{learning from human knowledge during the discussion},” which could “\emph{enhance the quality of decision-making}'' (3 participants). However, three participants found the lack of transparency in the update process led them to ``\emph{doubt the reliability of the AI’s initial opinions}.''

Although the AI's dynamic updates mimic human deliberation, making these updates more meaningful requires considering diverse user perceptions and designing the feature more thoughtfully. Specifically, the future design of AI opinion updates should draw insights from social science \cite{liao2021human}, addressing user preferences regarding the frequency and magnitude of updates. \ms{Recall that in the introduction before the experiment, we did not disclose the specific mechanism of AI opinion updates to the users, so this process was not transparent to them. Enhancing transparency and users' understanding of AI update could alleviate their concerns}.

\subsubsection{Opportunities for Future Improvements of \emph{Deliberative AI}}
Participants provided several suggestions for improving \emph{Deliberative AI} in future versions. Fifteen participants emphasized \textbf{the need for AI to develop a deeper understanding of human thoughts and arguments}, calling for more nuanced and context-aware communication. Eight participants stressed the importance of AI remembering previous interactions to enable more personalized and coherent conversations, avoiding repetition and improving the AI's responsiveness to specific perspectives.

Twelve participants suggested that \textbf{AI should move beyond statistical reasoning} to offer broader, more holistic solutions, while 10 participants advocated for deeper analysis and consideration of broader implications in each context. Transparency was also highlighted as crucial, with eight participants asking for clearer explanations about the AI’s changing opinions to foster trust and understanding.

Regarding the design of human-AI discussions, eight participants preferred a more conversational and human-like interaction, while 11 favored concise AI responses. Two participants recommended gradually disclosing the AI’s arguments, and three wanted more user control over steering the discussion. Additional suggestions included improving dialogue generation speed and enhancing the display of visual information beyond just text.


\section{Discussion}

In this paper, we introduce the \emph{Human-AI Deliberation} approach to address two key challenges in AI-assisted decision-making: insufficient analytical engagement with AI recommendations and limited support for resolving human-AI disagreements. Traditional interfaces often restrict users to accepting or rejecting AI suggestions as a whole, limiting nuanced understanding and collaboration. Our exploratory assessment demonstrates that \emph{Human-AI Deliberation} improves decision accuracy and fosters appropriate reliance on AI compared to conventional explainable AI systems. In this section, we discuss key implications, generalizability, limitations, and directions for future work.

\subsection{Deliberation as a New Paradigm Complementing Existing (X)AI Assistance}
The \emph{Human-AI Deliberation} approach introduces a conflict-driven discussion model that complements traditional AI assistance. \emph{Deliberative AI} builds on explainable AI (XAI) principles, challenging human perspectives while respecting their agency as decision-makers. It enhances AI assistants by fostering deeper analytical thinking, encouraging users to form independent opinions before seeing AI suggestions (as in the Cognitive Forcing Function approach \cite{buccinca2021trust}) and stimulating critical thinking through AI-generated questions (similar to \cite{danry2023don}). Additionally, \emph{Human-AI Deliberation} involves users in active discussions, improving communication and transparency with AI. This engagement helps lead to more informed and nuanced decisions.

\ms{
\textbf{Application Value of Deliberative AI}. Deliberative AI is well-suited for critical decision-making tasks, such as investment or hiring, where decision quality outweighs the need for speed or convenience. While engaging with Deliberative AI may demand more time and effort, decision-makers appreciate its ability to support well-considered outcomes. Additionally, Deliberative AI fosters user reflection by encouraging individuals to examine biases in their reasoning and evaluate differences between their perspectives and the AI's suggestions. Participants frequently emphasized this in their qualitative feedback, highlighting its value even in subjective decision contexts. Beyond enhancing objective accuracy, Deliberative AI promotes thoughtful consideration and introspection, which are vital for decisions requiring nuanced judgment.
}

\subsection{Reducing Human Over-Reliance by Exposing AI Mistakes}

\emph{Human-AI Deliberation} significantly reduces participants' tendency to over-rely on inaccurate AI suggestions. This approach utilizes cognitive forcing theory, which encourages forming independent opinions before viewing AI recommendations. This helps counteract \emph{anchoring bias}—the undue influence of initial AI predictions on subsequent judgments—and fosters more analytical, System 2 thinking ~\cite{kahneman2011thinking}. While our \emph{Explainable AI (XAI)} baseline also promotes independent opinion formation, \emph{Deliberative AI} proves more effective by involving deeper deliberative discussions, especially when facing conflicting viewpoints.

In the \emph{Deliberative AI} architecture, participants become more aware of AI's limitations through engagement with conflicting opinions, reducing over-reliance. Notably, 31\% of conversations involve participants questioning AI's logic, reflecting a critical evaluation of its insights. This approach effectively minimizes over-reliance without increasing under-reliance, as participants adjust their judgments based on deliberative dialogue. We recommend that AI-assisted decision-making systems should emphasize transparency by \textbf{explicitly} highlighting potential AI errors rather than merely suggesting that ``AI may make errors''.

\subsection{Human-AI Conflict Resolution: Key to Enhancing Collaboration}
We propose that addressing conflicts in decision-making is more beneficial than merely seeking consensus. Our approach, \emph{Human-AI Deliberation}, prioritizes resolving disagreements between humans and AI—an often overlooked aspect in current AI-assisted decision-making. Conflicts serve as a lens to uncover underlying flaws and biases, making them essential to effective human-AI collaboration. This conflict-centered methodology offers several advantages: it enhances decision-making accuracy by mitigating over-reliance on AI, fosters deeper introspection, and facilitates reconciliation between differing human perspectives and AI-generated recommendations. Additionally, addressing conflicts allows for the identification of biases in AI interpretations, thereby promoting fairness in decision-making \cite{kramer1990pretrial, hochman2015fairness}.

However, prioritizing conflict resolution can influence user experience. \ms{Our experimental results show that while \emph{Deliberative AI} demonstrates no significant differences from \emph{XAI} and \emph{Human Alone} in terms of mental demand, effort, or perceived complexity, it introduces a more complex decision-making process, resulting in lower user satisfaction. We identify three primary sources of the system's burden and propose solutions to mitigate them:
\begin{itemize}
    \item \textbf{Articulating Opinions Across Dimensions:} Currently, users must articulate their views comprehensively across multiple dimensions. Future interfaces could streamline this by letting users focus on dimensions they consider most critical or provide approximate opinions in natural language. AI could then identify the direction and intensity of human opinions from languages, allowing users to refine these interpretations as needed.
    \item \textbf{Dialogue Effort:} The AI currently initiates discussions on dimensions with significant disagreements. Future designs could let users take a more active role or adopt a bidirectional approach to enhance autonomy. Alternatively, AI could prompt users to reflect on specific dimensions and acknowledge the completion of reflection with simple actions, such as clicking an ``OK'' button, reducing dialogue input.
    \item \textbf{Emotional Challenges in Conflict Resolution:} Resolving conflicts with AI can be difficult and reduce satisfaction with \emph{Deliberative AI}. To improve this, interfaces could reframe conflicts as opportunities for reflection and feedback, using Constructive Conflict Theory \cite{kriesberg2007constructive} to emphasize that AI and humans share the same goals. Drawing from Positive Psychology \cite{noor2008positive}, AI could demonstrate active listening, show appreciation for user viewpoints, and foster a supportive atmosphere to reduce dissatisfaction and encourage deeper engagement.
\end{itemize}
}

\subsection{Obstacles to Discussion: Humans and AI Think Differently}

\textbf{Humans Use Heuristics and Logic, While AI Relies on Data:} Integrating LLMs and DS-Models enhances \emph{Deliberative AI} for dynamic human-AI discussions. However, AI’s data-centric approach often clashes with human decision-making, which relies on personal understanding, logic, heuristics, and creativity \cite{ghattas2014improving, evans2002logic, doswell2004weather, simon1990bounded, simon1997models}. We propose using DS-Models to guide LLMs on data-related discussions, while LLMs handle non-data matters autonomously. Despite this, engaging users with strong subjective perspectives remains challenging \cite{liu2023evaluating}.

\textbf{Existing Explanation Methods Fall Short:} Current XAI methods, such as local feature importance \cite{liao2021human}, inadequately explain specific feature contributions. Our approach, which provides data-related insights like distribution and value comparisons, aims to support feature-level discussions. Yet, challenges persist: data patterns may not match AI explanations precisely, and users often rely on subjective reasoning \cite{doswell2004weather, ghattas2014improving}. As Miller et al. note, "probabilities are not as important as causal links" \cite{miller2019explanation}. Thus, data-driven insights alone may not align with users’ cognitive processes or replicate human-like conversations.

Future human-AI deliberation designs should address these challenges by aligning with human intuition and cognitive processes, and by crafting AI explanations that facilitate more effective discussions.

\subsection{Ethical Concerns in Deliberative AI Design}

\textbf{Responsible Use of LLMs in Assisting Human Decision-Making:} LLMs are evolving rapidly but are prone to “hallucinations,” where they generate plausible but incorrect information \cite{ji2023survey}. Relying solely on LLM-generated opinions without appropriate fine-tuning is irresponsible. Even with fine-tuning, LLMs may produce unpredictable responses. We advocate for a responsible approach that integrates LLMs with DS-Models, where DS-Models guide LLM responses, positioning LLMs primarily as intermediaries between users and DS-Models. Although this approach limits LLM flexibility, it enhances security and control. Nonetheless, even with DS-Model guidance, inaccuracies can still occur. Researchers should exercise caution and transparency, informing users about the potential for errors and the limitations of LLM-generated information.

\textbf{Ethical Issues with AI Opinion Updates:} Research indicates users value AI’s responsiveness to their arguments \cite{zheng2023competent}. Our AI opinion update mechanism, which adjusts AI stances based on user input and prediction uncertainty, aims to reflect this need. While users appreciate feeling heard, ethical concerns arise, particularly about accountability. A key issue is who should be responsible if the AI, initially correct, updates to an incorrect prediction after discussion. Additionally, the AI’s adaptability may create the impression of learning and progress, even if the underlying model remains unchanged, potentially leading to unrealistic expectations. It is crucial to design AI systems with transparent updating mechanisms to ensure users understand how updates are made and manage their expectations effectively.

\subsection{On the Generalizability of Human-AI Deliberation}

\textbf{Task Suitability}: \emph{Human-AI Deliberation} is less suited for repetitive, low-stakes tasks like content moderation \cite{lai2022human} but is more appropriate for high-stakes, complex tasks such as healthcare \cite{lee2021human}, finance \cite{green2019principles}, and criminal justice \cite{chiang2023two}.

\textbf{Discussion Effort}: Discussing every feature, as in our study, may be impractical for tasks with many attributes. Grouping features into broader categories could streamline discussions.

\textbf{Opinion Representation and Alignment}: We used the Weight of Evidence method to quantify opinions, but this may not be intuitive for all users. Future designs could infer opinions from natural language, emotional intensity, or comparisons, or simplify input with rankings or pairwise comparisons \cite{furnkranz2010preference}.

\textbf{Data Type Applicability}: While designed for tabular data, the architecture can extend to text tasks with LLM's capability in dealing with textual data. Adapting it to image data may require advances in vision-language models \cite{zhang2024vision}.

\ms{\textbf{Using LLM as a Communication Bridge}: In Deliberative AI, the LLM acts as a deliberation facilitator, intention analyzer, and argument evaluator, relying mainly on its language understanding capabilities with minimal reliance on its reasoning abilities. Through two pilot studies, we validated the LLM's effectiveness in these roles. However, challenges remain regarding its reasoning capabilities \cite{wei2022chain}. Designers should leverage the LLM's strengths in communication and avoid overburdening it with complex reasoning tasks.}

\subsection{Limitations and Future Work}

The study design has several limitations. \textbf{First}, the college admissions task used for illustration does not fully capture real-world admissions processes, which typically involve in-depth discussions about a student's materials, such as her/his statement of purpose (SOP), background, and overall fit for the department. However, our used public dataset quantifies the ``strength'' of the SOP and recommendation letter as a scale value, which inevitably oversimplifies these nuanced evaluations. Additionally, while participants had relevant experience, they were not actual admissions committee members, leading to an expertise gap. Furthermore, most participants had experience applying to master's programs, with only a few familiar with PhD admissions. Given the distinct evaluation criteria for these two application types, our findings may not generalize to PhD admissions. Future research should assess the approach’s effectiveness in real-world admissions contexts. \textbf{Second}, graduate admissions is a subjective task that lacks a definitive ground truth. In our study, since the dataset labels were provided by a professional admissions committee, we used decision accuracy and over/under-reliance as objective metrics to assess decision quality ``to some extent''. The dataset also abstracts subjective elements (e.g., SOP and recommendation letter) into numerical strength values, making the task relatively more objective. Nonetheless, future work should explore the effects of Deliberative AI in the context of more objective tasks. \textbf{Third}, the number of decision tasks in the study was limited. During the pilot, using 8-10 tasks led to participant fatigue and a drop in engagement after completing 3-4 tasks. To maintain engagement, we reduced the task count to four, which restricts the generalizability of our results. Future studies should conduct long-term evaluations to collect more deliberative decision data. \textbf{Fourth}, the independent decision accuracy of both humans and AI was relatively low (50-60\%) in our selected task cases, as we intentionally chose tasks prone to conflicts that require discussion. The ambiguity in these cases led to suboptimal performance from both. However, our study did not address less-ambiguous cases, where differing but firm opinions may still arise. We believe Deliberative AI can still help reduce errors in such cases by prompting reflection on biases or overlooked perspectives, especially when humans' confidence is high. Further research is needed to explore different task cases for a more comprehensive understanding of human-AI deliberation.

\section{Conclusion}
In this paper, we introduce \emph{Human-AI Deliberation}, as a new paradigm of AI assistance for decision-making. \emph{Human-AI Deliberation} encourages the externalization of thoughts, facilitates interactive deliberation between humans and AI, and allows for dynamic updates of decisions. To enable the deliberation, we present a novel AI assistant called \emph{Deliberative AI}, which can identify differences in viewpoints, engage in comprehensive deliberation, and adapt its suggestions during discussions. We apply this architecture to an illustrative task (graduate admissions decisions) and conduct an exploratory study to assess its potential impact on decision-making processes, outcomes, user perceptions, and experiences. Results indicate the potential of \emph{Deliberative AI} to improve decision accuracy and promote more appropriate human reliance on AI. Additionally, we analyze participants' open-ended feedback to gain deeper insights into how users use and perceive \emph{Deliberative AI}, uncovering areas for improvement. With the key insights and implications derived from our study, we aim for this work to serve as an exploratory step toward establishing a new paradigm of AI assistance that enhances decision-making.

\begin{acks}
This work is supported by the Research Grants Council of the Hong Kong Special Administrative Region under General Research Fund (GRF) with Grant No. 16207923.
\end{acks}

\bibliographystyle{ACM-Reference-Format}
\bibliography{sample-base}

\appendix
\section*{Appendices}
\section{Baseline XAI Interface}

\begin{figure*}[htbp]
	\centering 
	\includegraphics[width=\linewidth]{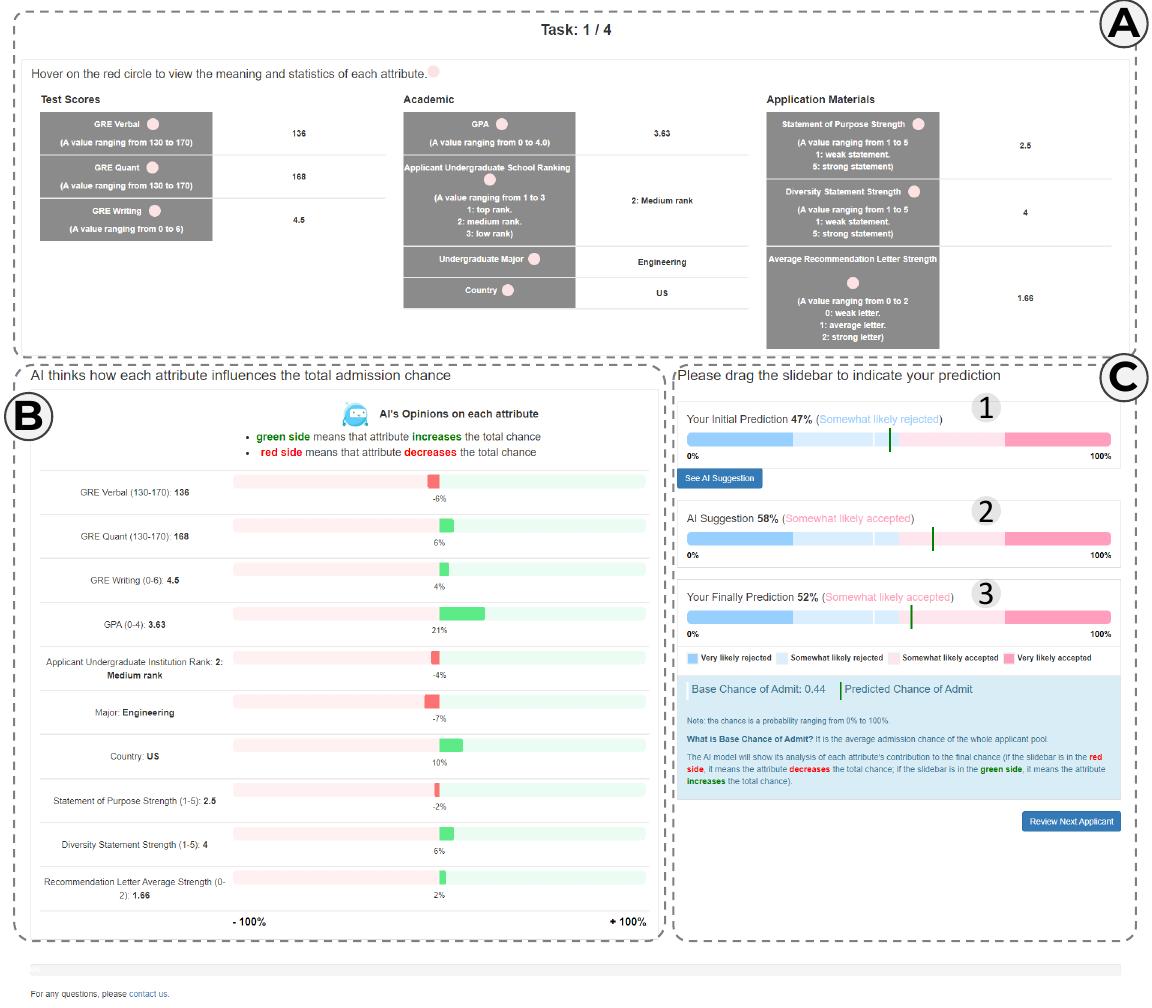}
	\caption{The baseline XAI (traditional explainable AI) interface in our user study. The interface contains three parts. The top (A) is the applicant's profile. The bottom left part (B) shows AI's feature contribution explanation. The bottom right part (C) is for humans to (1) indicate their initial predictions, (2) see AI's suggestions, and (3) indicate their final decisions. Note that AI's suggestions and explanations are only shown after humans make their initial predictions. (All the dashed lines are only for illustration)}
	\label{fig:baselineinterface}
        \Description{}
\end{figure*}

Figure \ref{fig:baselineinterface} shows the baseline XAI interface.

\section{An Example Dataflow of Deliberative AI}

\begin{figure*}[htbp]
	\centering 
	\includegraphics[width=\linewidth]{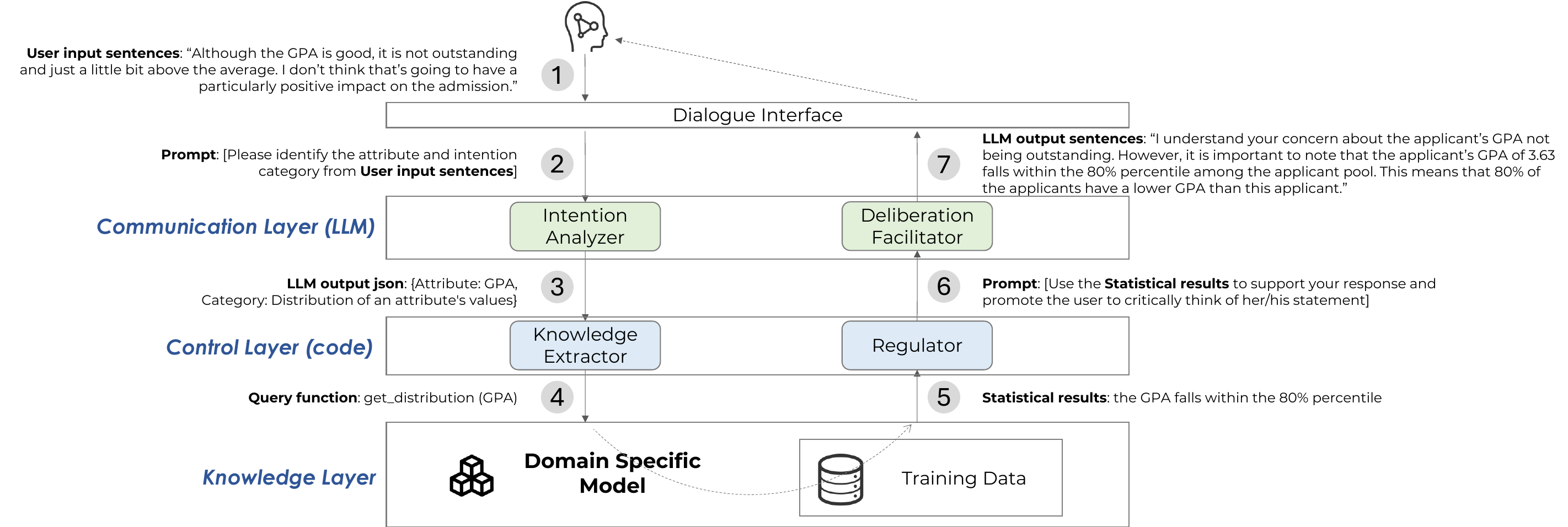}
	\caption{An illustration of how \emph{Deliberative AI} processes humans' inputs and how it generates outputs. The prompts used are simplified in this figure for illustration purposes. (for the complete prompts, please check our supplementary materials)}
	\label{fig:io}
        \Description{}
\end{figure*}

Figure \ref{fig:io} provides the details of a conversation between a human and a \emph{Deliberative AI} discussing how an applicant's GPA affects admissions chances. Here's a step-by-step breakdown:
\begin{enumerate}
    \item The user inputs GPA-related arguments in the dialogue interface.
    \item The system packages the user's input as a prompt for the \emph{Intention Analyzer} in the Communication Layer.
    \item The \emph{Intention Analyzer} recognizes attributes and intentions and saves in JSON format, then forwards it to the \emph{Knowledge Extractor} in the Control Layer.
    \item The \emph{Knowledge Extractor} generates a query function and fetches statistical results from the DS-Model and training data.
    \item The statistical results are transmitted to the \emph{Regulator}.
    \item \emph{Regulator} crafts a constraint prompt ensuring consistency between the LLM's output and the DS-Model's prediction, feeding it to the LLM-based \emph{Deliberation Facilitator}.
    \item The \emph{Deliberation Facilitator} generates responses to the user's initial arguments.
\end{enumerate}


\section{Detailed Metrics and Questions}
Table \ref{tab:measurement} shows the detailed metrics and questions used in our measurement.

\renewcommand{\arraystretch}{1.5}
\begin{table*}[htp]  

\centering  
\fontsize{8}{8}\selectfont  

\caption{Measurements used in our user study. We collected participants' objective decision data, subjective questionnaire data, and qualitative open-ended feedback.}\label{tab:measurement}

\begin{tabular}{m{1.5cm}<{\centering}m{3.5cm}<{\centering}m{9cm}}
\toprule
\textbf{Aspect}&\textbf{Metrics}&\textbf{Detailed Meaning and Questions}\\ \hline
\multicolumn{3}{l}{\textbf{Objective Measures}}\\
\hline
\multirow{1}*{\shortstack{Performance}}&Decision Accuracy&Accuracy of participants' final predictions.\\ 

\hline

\multirow{7}*{\shortstack{Reliance}}&Agreement Fraction&Percentage of tasks where participants’ final prediction agreed with AI’s prediction. $\rm \frac{Number\ of\ final\ decisions\ same\ as\ the\ AI\ suggestion}{Total\ number\ of\ decisions}$\\
\cline{2-3}
&Switch Fraction&Percentage of tasks where AI's prediction was used when initial disagreement existed. $\rm \frac{Number\ of\ decisions\ user\ switched\ to\ agree\ with\ the\ AI\ model}{Total\ number\ of\ decisions\ with\ initial\ disagreement}$\\ 
\cline{2-3}
&Over-reliance Ratio&Fraction of tasks where participants used an incorrect AI prediction. $\rm \frac{Number\ of\ incorrect\ human\ final\ decisions\ with\ incorrect\ AI\ suggestions}{Total\ number\ of\ incorrect\ AI\ suggestions}$\\ 
\cline{2-3}
&Under-reliance Ratio&Fraction of tasks where participants did not use a correct AI prediction. $\rm \frac{Number\ of\ incorrect\ human\ final\ decisions\ with\ correct\ AI\ suggestions}{Total\ number\ of\ correct\ AI\ suggestions}$\\ 
\hline
 
\multicolumn{3}{l}{\textbf{Subjective Measures}}\\
\hline

\multirow{4}*{\shortstack{Perceptions\\of AI}}&Helpfulness&\emph{``I think the AI model's assistance is helpful/useful for me to make good decisions.''}\\
\cline{2-3}
&Trustworthiness&\emph{``The AI model can be trusted to provide reliable decision support.''}\\
\cline{2-3}
&Understanding&\emph{``I understand how the AI model works to predict an applicant's chance of being admitted.''}\\
\hline

\multirow{6}*{\shortstack{User\\Experience}}&Decision Confidence&\emph{``I feel confident in the decisions I made.''}\\
\cline{2-3}
&Mental Demand&\emph{``The decision-making process is mentally demanding.''}\\
\cline{2-3}
&Effort&\emph{``I have to work hard (mentally and physically) to accomplish my level of performance.''}\\ 
\cline{2-3}
&Complexity&\emph{``The decision-making process and the interaction with AI models are complex.''}\\ 
\cline{2-3}
&Satisfaction&\emph{``I am satisfied with the AI model's assistance and the decision-making process.''}\\
\hline

\multirow{4}*{\shortstack{Open-ended\\Feedback}}&Perception of helpfulness&\emph{``Do you think the discussion with AI is (or not) helpful? Could you tell us the reasons why you think the discussion is helpful (or not helpful)?''}\\
\cline{2-3}
&Perception of AI update&\emph{``What do you think of the AI updating its own views during the discussion?''}\\
\cline{2-3}
&Potential Improvement&\emph{``To make a better discussion, which parts do you think the current AI needs to be improved, and how should it be improved?''}\\
\bottomrule
\end{tabular}
\end{table*}

\end{document}